\documentclass[sigconf]{acmart}

\copyrightyear{2023}
\acmYear{2023}
\setcopyright{none}
\acmConference[WSDM '23]{Proceedings of the Sixteenth ACM International Conference on Web Search and Data Mining}{February 27-March 3, 2023}{Singapore, Singapore} 
\acmBooktitle{Proceedings of the Sixteenth ACM International Conference on Web Search and Data Mining (WSDM '23), February 27-March 3, 2023, Singapore, Singapore}
\acmPrice{15.00}
\acmDOI{10.1145/3539597.3570427}
\acmISBN{978-1-4503-9407-9/23/02}

\usepackage{balance}

\usepackage{hyperref}
\makeatletter
\g@addto@macro{\UrlBreaks}{\UrlOrds}
\makeatother

\usepackage{hyperref}
\usepackage{amsmath}
\usepackage{amsfonts}

\newtheorem{myprob}{Problem}

\usepackage{bbm}
\usepackage{dsfont}

\usepackage{algorithm}[h]
\usepackage[noend]{algpseudocode}

\usepackage{paralist}
\usepackage{booktabs} 
\usepackage{multirow}

\usepackage{subcaption}
\usepackage{epstopdf}
\usepackage{graphics}
\DeclareGraphicsExtensions{.eps,.gz,.eps,.pdf,.png,.jpg}
\graphicspath{{./}{./fig/}}

\usepackage{qtree} 
\usepackage{graphicx}

\newcommand{\sstitle}[1]{\noindent\textbf{#1.\/}}

\def\Snospace~{\S{}}

\makeatletter
\newcommand{\removelatexerror}{\let\@latex@error\@gobble}
\makeatother

\begin{document}

\setlength{\belowdisplayskip}{3pt}
\setlength{\belowdisplayshortskip}{3pt}
\setlength{\abovedisplayskip}{3pt}
\setlength{\abovedisplayshortskip}{3pt}









\title[Joint Stock Movement Prediction with Multi-Order Deep Learning]{Efficient Integration of Multi-Order Dynamics and Internal Dynamics in 
Stock Movement Prediction}

\author{Thanh Trung Huynh}
\orcid{0000-0003-2027-5362}
\affiliation{%
  \institution{Ecole Polytechnique Federale de Lausanne, Switzerland}
}

\author{Minh Hieu Nguyen}
\orcid{0000-0003-1518-8977}
\affiliation{%
  \institution{Hanoi University of Science and Technology, Vietnam}
}

\author{Thanh Tam Nguyen}
\authornote{Corresponding author.}
\orcid{0000-0002-2586-7757}
\affiliation{%
  \institution{Griffith University, Australia}
}

\author{Phi Le Nguyen}
\orcid{0000-0001-6547-7641}
\affiliation{%
  \institution{Hanoi University of Science and Technology, Vietnam}
}

\author{Matthias Weidlich}
\orcid{0000-0003-3325-7227}
\affiliation{%
  \institution{Humboldt-Universit\"at zu Berlin, Germany}
}

\author{Quoc Viet Hung Nguyen}
\orcid{0000-0002-9687-1315}
\affiliation{%
  \institution{Griffith University, Australia}
}

\author{Karl Aberer}
\orcid{0000-0003-3005-7342}
\affiliation{%
  \institution{Ecole Polytechnique Federale de Lausanne, Switzerland}
}


\renewcommand{\shortauthors}{Huynh et al.}

\begin{abstract}

Advances in deep neural network (DNN) architectures have enabled new prediction 
techniques for stock market data.
Unlike other multivariate time-series data, stock markets show two unique 
characteristics: (i) \emph{multi-order dynamics}, as stock prices are affected 
by strong non-pairwise correlations (e.g., within the same industry); and (ii) 
\emph{internal dynamics}, as each individual stock shows some particular 
behaviour. 
Recent DNN-based methods capture multi-order dynamics using hypergraphs, but 
rely on the Fourier basis in the convolution, which is both inefficient and 
ineffective. In addition, they largely ignore internal dynamics by adopting the 
same model for each stock, which implies a severe information loss. 


In this paper, we propose a framework for stock movement prediction to overcome 
the above 
issues. Specifically, the framework includes temporal generative filters that 
implement a memory-based mechanism onto an LSTM network in an attempt to learn 
individual patterns per stock. 
Moreover, we employ hypergraph attentions to capture the non-pairwise 
correlations. Here, using the wavelet basis instead of the Fourier basis, 
enables us to simplify the message passing and focus on the localized 
convolution. Experiments with US market data over six years show that our 
framework outperforms state-of-the-art methods in terms of profit and stability.
Our source code and data are available at \url{https://github.com/thanhtrunghuynh93/estimate}. 
\end{abstract}

\ccsdesc[500]{Computing methodologies~Neural networks}

\keywords{hypergraph embedding, stock market, temporal generative filters} 

\maketitle

\section{Introduction}
\label{sec:intro}

The stock market denotes a financial ecosystem where the stock shares that represents the ownership of businesses are held and traded among the investors, 
with a market capitalization of more than 93.7\$ trillion globally at the end 
of 2020~\cite{wang2021coupling}. In recent years, approaches for automated 
trading emerged that are driven by artificial 
intelligence (AI) models. They continuously analyze the market behaviour and 
predict the short-term trends in stock prices. While these methods struggle 
to understand the complex rationales behind such trends (e.g., macroeconomic 
factors, crowd behaviour, and companies' intrinsic values), they have been 
shown to yield accurate predictions. Moreover, they track market changes in 
real-time, by observing massive volumes of trading data and indicators, and 
hence, enable quick responses to events, such as a market crash. Also, they are 
relatively robust against emotional effects (greed, fear) that tend to 
influence human traders~\cite{nourbakhsh2020spread}. 

Stock market analysis has received much attention in the past. Early work relies on handcrafted features, a.k.a technical indicators, to model the stock movement. For example, RIMA~\cite{piccolo1990distance}, a popular time-series statistics model, may be applied to moving averages of stock prices to derive price predictions~\cite{ariyo2014stock}. However, handcrafted features tend to lag behind the actual price movements. 
Therefore, recent approaches adopt deep learning to model the market based on historic data. Specifically, recurrent neural networks (RNN)~\cite{chen2015lstm} have been employed to learn temporal patterns from the historic data and, based thereon, efficiently derive short-term price predictions using regression~\cite{nelson2017stock} or classification~\cite{zhang2017stock}. 

\begin{figure}[ht]
	\centering
	\vspace{-1em}
    \includegraphics[width=1\linewidth]{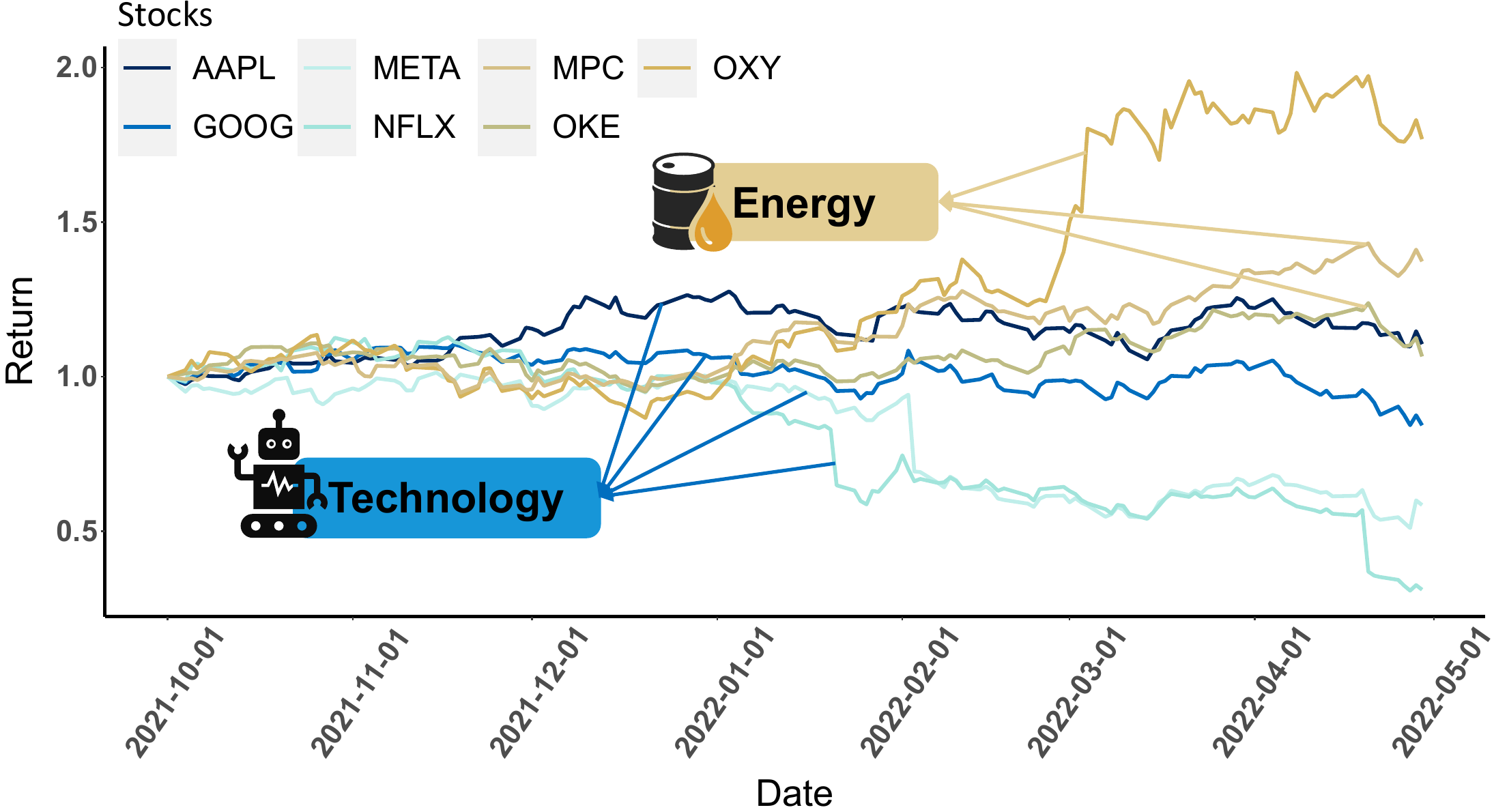}
    \vspace{-2em}
    \caption{Illustration of complex stock price correlation}
    \label{fig:quant_example}
    \vspace{-1em}
\end{figure}

However, stock market analysis based on deep learning faces two important requirements. 
First, multi-order dynamics of stock movements need to be incorporated. Price movements are often correlated 
within a specific group of stocks, e.g., companies of the same industry sector that are affected by the same government policies, laws, and tax rates. For instance, as shown in \autoref{fig:quant_example}, in early 2022, prices for US technology stocks (APPL (Apple), META (Facebook), GOOG (Google), NFLX (Netflix)) went down due to the general economic trend (inflation, increased interest rates), whereas stocks in the energy sector, like MPC, OKE, or OXY, experienced upward trends due to oil shortages caused by the Russia-Ukraine war. Second, the internal dynamics per stock need to be incorporated. In practice, even when considering highly correlated stocks, there is commonly still some individual behaviour. For example, in \autoref{fig:quant_example}, APPL and GOOG stocks 
decrease less severely than META and NFLX, as the former companies (Apple, Google) maintain a wider and more sustainable portfolio compared to the latter two (Facebook, Netflix)~\cite{nytimes-stock}. 

Existing work provides only limited support for these requirements. First, to incorporate multi-order dynamics of stock markets, RNNs can be combined with 
graph neural networks (GNNs)~\cite{kim2019hats}. Here, state-of-the-art solutions adopt hypergraphs, in which an edge captures the correlation of multiple stocks~\cite{sawhney2020spatiotemporal, 	sawhney2021stock}. 
Yet, these approaches rely on the Fourier basis in the convolution, which implies costly matrix operations and does not maintain the localization well. This raises the question of how to achieve an efficient and effective convolution process for hypergraphs (\textbf{Challenge 1}). Moreover, state-of-the-art approaches apply a single RNN to all stocks, thereby 
ignoring their individual behaviour. The reason being that maintaining a separate model per stock would be intractable with existing techniques. This raises the question of how to model the internal dynamics of stocks efficiently (\textbf{Challenge 2}).

In this work, we address the above challenges by proposing \textbf{E}fficient 
\textbf{St}ock \textbf{I}ntegration with Te\textbf{m}poral Generative Filters 
and Wavelet Hypergraph \textbf{At}t\textbf{e}ntions (ESTIMATE), a profit-driven framework for quantitative trading. Based on the aforementioned 
idea of adopting hypergraphs to capture non-pair\-wise correlations between 
stocks, the framework includes two main contributions:
\begin{compactitem}
	\item We present temporal generative filters that implement a hybrid 
	attention-based LSTM architecture to capture the stocks' individual 
	behavioural patterns (\textbf{Challenge 2}). These patterns are then 
	fed to hypergraph convolution layers to obtain spatio-temporal embeddings 
	that are optimized with respect to 
	the potential of the stocks for short-term profit. 
	\item We propose a mechanism that combines the temporal patterns of stocks 
	with spatial convolutions through hypergraph attention, thereby integrating 
	the internal dynamics and the multi-order dynamics. Our convolution process 
	uses the wavelet basis, which is efficient and also effective in terms of 
	maintaining the localization (\textbf{Challenge 1}). 
\end{compactitem}
To evaluate our approach, we report on backtesting experiments for the US 
market. Here, we try to simulate the 
real trading actions with a strategy for portfolio management and risk control. 
The results 
demonstrate the robustness of our technique compared to existing approaches in 
terms of stability and return. Our source code and data are 
available~\cite{sourcecode}.

The remainder of the paper is organised as follows. \autoref{sec:model} 
introduces the problem statement and gives an overview of our approach. We 
present our new techniques, the temporal generative filters and wavelet 
hypergraph attentions, in \autoref{sec:local_emb} and \autoref{sec:market_agg}. 
\autoref{sec:exp} presents experiments, \autoref{sec:related} reviews related 
works, and \autoref{sec:con} concludes the paper.

\section{Model and approach}
\label{sec:model}

\subsection{Problem Formulation}

In this section, we formulate the problem of predicting the trend of a stock in 
the short term. We start with some basic notions. 

\sstitle{OHCLV data} At timestep $t$, the open-high-low-close-volume (OHLCV) 
record for a stock $s$ is a vector $x_s^t = [o^t_s, h^t_s, l^t_s, c^t_s, 
v^t_s]$. It denotes the open, high, low, and close price, and the 
volume of shares that have been traded within that timestep, respectively.

\sstitle{Relative price change} We denote the relative close price change 
between two timesteps $t_1 < t_2$ of stock $s$ by $d^{(t_1, t_2)}_s = 
({c^{t_2}_s - c^{t_1}_s})/{c^{t_1}_s}$. The relative price change normalizes 
the market price variety between different stocks in comparison to 
the absolute price change. 

Following existing work on stock market 
analysis~\cite{sawhney2020spatiotemporal,alstm}, we focus on the prediction of 
the change in price rather than the absolute value. The reason being that the 
timeseries of stock prices are non-stationary, whereas their changes are 
stationary~\cite{li2020modeling}. 
Also, this avoids the problem that 
forecasts often lag behind the actual value~\cite{kim2019hats,hu2018listening}. 
We thus define the addressed problem as follows:


\begin{myprob}[Stock Movement Prediction]
\label{prob:movement}
Given a set $S$ of stocks and a lookback window of $k$ trading days of historic 
OHLCV  records $x_s^{(t-k-1) \dots t}$ for each stock $s\in S$, the problem of 
\emph{Stock Movement Prediction} is to predict the relative price change 
$d^{(t, t + w)}_s$ for each stock in a short-term lookahead window $w$.
\end{myprob}
We formulate the problem as a short-term regression for several reasons. 
First, we consider a lookahead window over next-day prediction to be robust 
against random market fluctuations~\cite{zhang2017stock}. Second, we opt for 
short-term prediction, as an estimation of the long-term trend is commonly 
considered infeasible without the integration of expert knowledge on the 
intrinsic value of companies and on macroeconomic effects. 
Third, we focus on a regression problem instead of a classification problem to 
incorporate the magnitude of a stock's trend, which is important for 
interpretation~\cite{gu2020price}.


\begin{figure*}[ht]
	\centering
	\includegraphics[width=\linewidth]{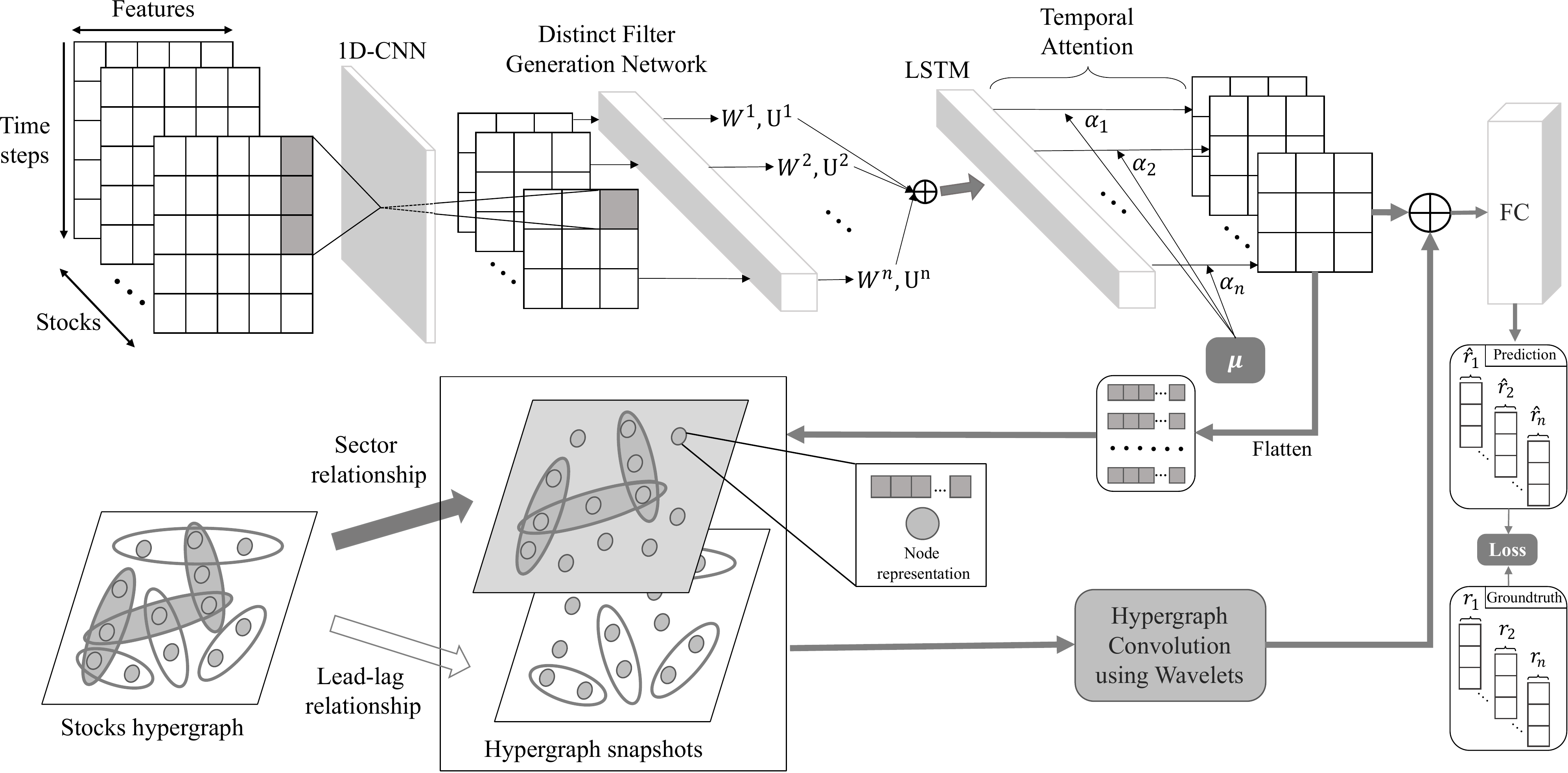}	
	\caption{Overview of our framework for stock movement prediction.}
	\label{fig:framework}	
\end{figure*}

\subsection{Design Principles}

We argue that any solution to the above problem shall satisfy the following 
requirements:

\begin{itemize}
    \item \textbf{R1: Multi-dimensional data integration:} Stock market data is 
    multivariate, covering multiple stocks and multiple features 
    per stock. A solution shall integrate these data dimensions and support the 
    construction of additional indicators from basic OHCLV data. 
    
    \item \textbf{R2: Non-stationary awareness:} The stock market is driven by 
    various factors, such as socio-economic effects or supply-demand changes. 
    Therefore, a solution shall be robust against non-predictable behaviour of 
    the market. 
    
    \item \textbf{R3: Analysis of multi-order dynamics:} The relations between 
    stocks are complex (e.g., companies may both, cooperate and compete) and 
    may evolve 
    over time. A solution thus needs to analyse the multi-order 
    dynamics in a market. 
    
    \item \textbf{R4: Analysis of internal dynamics:} Each stock also shows 
    some individual behaviour, beyond the multi-order correlations induced by 
    market segments. A solution therefore needs to analyse and integrate such 
    behaviour for each stock. 

\end{itemize}

\subsection{Approach Overview}

To address the problem of stock movement prediction in the light of the above 
design principles, we propose the framework shown in \autoref{fig:framework}.
It takes historic data in the form of OHCLV records and derives a model for 
short-term prediction of price changes per stock. 

Our framework incorporates requirement 
\textbf{R1} by first extracting the historic patterns per stock using 
a temporal attention LSTM. Here, the attention mechanism is used along with 
a 1D-CNN to assess the impact of the previous timesteps. In addition to the 
OHCLV data, we employ technical indicators to mitigate the impact of noisy 
market behaviour, thereby addressing requirement \textbf{R2}.
Moreover, we go beyond the state of the art by associating the core 
LSTM parameters with a learnable vector for each stock. It serves as a memory 
that stores its individual information (requirement \textbf{R4}) and results in 
a system of temporal generative filters. We explain the details of these 
filters in \autoref{sec:local_emb}. 

To handle multi-order dynamics (requirement \textbf{R3}), we model the market 
with an industry-based hypergraph, which naturally presents non-pairwise 
relationships. We then develop a wavelet convolution mechanism, which leverages 
the wavelet basis to achieve a simpler convolution process than existing 
approaches. We apply a regression loss to steer the model to predict the 
short-term trend of each stock price. The details of our proposed hypergraph 
convolution process are given in \autoref{sec:market_agg}.

\section{Temporal Generative Filters}
\label{sec:local_emb}

This section describes our temporal generative filters used to capture the 
internal dynamics of stocks. 

\sstitle{Technical indicators} We first compute various technical indicators 
from the input data in order to enrich the data and capture the historical 
context of each stock. These indicators, summarized in \autoref{tab:notation}, 
are widely used in finance. 
For each stock, we concatenate these indicators to form a stock price 
feature vector $x_t$ on day $t$. This vector is then 
forwarded through a multi-layer perceptron (MLP) layer to modulate the input 
size. 
 
\begin{table}[!ht]
\centering
\scriptsize
\vspace{-1em}
\caption{Summary of technical indicators used.}     
\label{tab:notation}
\vspace{-1.5em}
\resizebox{\linewidth}{!}{%
\begin{tabular}{c l}
\toprule
\textbf{Type} & \textbf{Indicators} \\
\midrule
Trend Indicators & Arithmetic ratio, Close Ratio, Close SMA, \\ 
 & Volume SMA, Close EMA, Volume EMA, ADX \\
Oscillator Indicators & RSI, MACD, Stochastics, MFI \\ 
Volatility Indicators & ATR, Bollinger Band, OBV \\
\bottomrule
\end{tabular}
}
\vspace{-1em}
\end{table}

\sstitle{Local trends} To capture local trends in stock patterns, we employ 
convolutional neural networks (CNN). By compressing the length of the series of 
stock features, they help to mitigate the issue of long-term dependencies. As 
each feature is a one-dimensional timeseries, we apply one-dimensional filters 
(1D-CNN) over all timesteps: 
\begin{equation}
x^k_l = b^l_k + conv1D(w^{l-1}_{ik}, s^{l-1}_i)
\end{equation}
where $x^k_l$ represent the input feature at the $k^{th}$ neuron of layer $l$; 
$b^l_k$ is the corresponding bias; $w^{l-1}_{ik}$ is the kernel from the 
$i^{th}$ neuron at layer $l - 1$ to the $k^{th}$ neuron at layer $l$; and 
$s^{l-1}_i$ is the output of the $i^{th}$ neuron at layer $l - 1$.

\sstitle{Temporal LSTM extractor with Distinct Generative Filter} After 
forwarding the features through the CNNs, we use an LSTM to capture the 
temporal dependencies, exploiting its ability to memorize long-term 
information. Given the concatenated feature $q_t$ of the stocks at time $t$, we 
feed the feature through the LSTM layer: 
\begin{equation}
\label{eqn:LSTM}
h_k = LSTM(x_k , h_{k-1}), t - T \leq k \leq t - 1
\end{equation}
where $h_k \in \mathbb{R}^d$ is the hidden state for day $l$ and d is the 
hidden state dimension. The specific computation in each LSTM unit includes:
\begin{align*}
& i_t = \sigma(W_{xi}x_t + U_{hi}h_{t-1} +b_i), &
f_t = \sigma(W_{xf}x_t + U_{hf}h_{t-1} +b_f), \\
& g_t = \textit{tanh}(W_{xg}x_t + U_{hg}h_{t-1} +b_g), &
o_t = \sigma(W_{xo}x_t + U_{ho}h_{t-1} +b_o), \\
& c_t = f_t \odot c_{t-1} + i_t \odot g_t, &
h_t = o_t \odot \textit{tanh}(c_t).
\end{align*}

As mentioned, existing approaches apply the same LSTM to the 
historical data of different stocks, which results in the learned set 
of filters ($\mathbb{W} = \{W_{xi}, W_{xf}, W_{xg}, W_{xo}\}, \mathbb{U} = 
\{U_{xi}, U_{xf}, U_{xg}, U_{xo}\}$) representing the average temporal 
dynamics. This is insufficient to capture each 
stock's distinct behaviour (\textbf{Challenge 2}). 

A straightforward solution would be to 
learn and store a \emph{set} of LSTM filters, one for each stock. Yet, such an 
approach quickly becomes intractable, 
especially when the number 
of stocks is large. 

In our model, we overcome this issue by 
proposing a memory-based mechanism onto the LSTM network to learn 
the individual patterns per stock, while not expanding the core LSTM. 
Specifically, we first assign to each stock $i$ a memory $M^i$ in the form of a 
learnable m-dimensional vector, $M^i \in \mathbb{R}^m$. Then, for each entity, 
we feed the memory through a Distinct Generative Filter, denoted by 
\textit{DGF}, to obtain the weights ($\mathbb{W}^i, \mathbb{U}^i$) of 
the LSTM network for each stock:
\begin{equation}
\mathbb{W}^i, \mathbb{U}^i = DGF(M^i)
\end{equation}
Note that \textit{DGF} can be any neural network architecture, such as a CNN or 
an MLP. In our work, we choose a 2-layer MLP as \textit{DGF}, as it is simple 
yet effective. As the \textit{DGF} is required to generate a set of eight 
filters 
$\{W^i_{xi}, W^i_{xf}, W^i_{xg}, W^i_{xo}, U^i_{xi}, U^i_{xf}, U^i_{xg}, 
U^i_{xo}\}$ from $M^i$, we generate a concatenation of the filters and then 
obtain the results by splitting. Finally, replacing the common filters by the 
specific ones for each stock in \autoref{eqn:LSTM}, we have:
\begin{equation}
h^i_k = LSTM(x^i_k , h^i_{k-1} \mid \mathbb{W}^i, \mathbb{U}^i), t - T \leq k 
\leq t - 1
\end{equation}
where $h^i_k$ is the hidden feature of each stock $i$. 

To increase the efficiency of the LSTM, we apply a temporal attention 
mechanism to guide the learning process towards important 
historical features. The attention mechanism attempts to aggregate temporal 
hidden states $\hat{h}^i_k = [h^i_{t - T}, \dots, h^i_{t - 1}]$ from previous 
days into an overall representation using learned attention weights: 
\begin{equation}
	\label{eq:att_weights}
\mu(\hat{h}^i_t) = \sum_{k=(t - T)}^{t-1}{\alpha_k h^i_k} = \sum_{k=(t - T)}^{t-1}{\frac{\mathrm{exp}(h^{i^T}_k)W\hat{h}^i_t}{\sum_k{\mathrm{exp}(h^{i^T}_k)W\hat{h}^i_t}} h^i_k}
\end{equation}
where $W$ is a linear transformation, $\alpha_k = 
\frac{\mathrm{exp}(h^{i^T}_k)W\hat{h}^i_t}{\sum_k{\mathrm{exp}(h^{i^T}_k)W\hat{h}^i_t}}
 h^i_k$ are the attention weights using softmax. To handle the 
non-stationary nature of the stock market, we leverage the Hawkes 
process~\cite{bacry2015hawkes}, as suggested for financial 
timeseries in~\cite{sawhney2021stock}, to enhance the temporal 
attention mechanism in \autoref{eq:att_weights}. The Hawkes process is a 
``self-exciting'' temporal point process, where some random event ``excites'' 
the process and increases the chance of a subsequent other random event (e.g., 
a crises or policy change). To realize the Hawke process, the attention 
mechanism also learns an excitation parameter $\epsilon_k$ of the day $k$ and a 
corresponding decay parameter $\gamma$:
\begin{equation}
\hat{\mu}(\hat{h}^i_t) = \sum_{k=(t - T)}^{t-1}{\alpha_k h^i_k} + \epsilon_k \mathrm{max}(\alpha_k h^i_k, 0) \mathrm{exp}(-\gamma\alpha_k h^i_k)
\end{equation}
Finally, we concatenate the extracted temporal feature $z_i = 
\hat{\mu}(\hat{h}^i_t)$ of each stock to form $\mathbf{Z_T} \in \mathbb{R}^{n 
\times d}$, where $n$ is the number of stocks and $d$ is the 
embedding dimension.

\section{High-order market learning with wavelet hypergraph attentions} 
\label{sec:market_agg}

To model the groupwise relations between stocks, we aggregate the learned 
temporal patterns of each stock over a hypergraph that represents multi-order 
relations of the market. 

\sstitle{Industry hypergraph} To model the interdependence between stocks, we 
first initialize a hypergraph based on the industry of the respective 
companies. Mathematically, the industry hypergraph is denoted as $\mathbb{G}_i 
= (S, E_i, w_i)$, where $S$ is the set of stocks and $E_i$ is the set of 
hyperedges; each hyperedge $e_i \in E_i$ connects the stocks that belong to the 
same industry. The hyperedge $e_i$ is also assigned a weight $w_i$ that 
reflects the importance of the industry, which we derive from the market 
capital of all related stocks.

\sstitle{Price correlation augmentation} Following the Efficient Market 
Hypothesis~\cite{malkiel1989efficient}, fundamentally correlated stocks 
maintain similar price patterns, which can be used to reveal the missing 
endogenous relations in addition to the industry assignment. To this end, for 
the start of each training and testing period, we calculate the price 
correlation between the stocks using the historical price of the last 1-year 
period. We employ the lead-lag correlation and the clustering method proposed 
in~\cite{bennett2021detection} to simulate the lag of the stock market, where a 
leading stock affects the trend of the rests. Then, we form hyperedges from the 
resulting clusters and add them to $E_i$. The hyperedge weight is, again, 
derived from the total market capital of the related stocks. We denote the 
augmented hypergraph by $\mathbb{G} = (\textbf{A}, \textbf{W})$, with 
$\textbf{A}$ and $\textbf{W}$ being the hypergraph incidence matrix and the 
hyperedge weights, respectively.

\begin{figure*}[!h]
    \centering
    \includegraphics[width=1\linewidth]{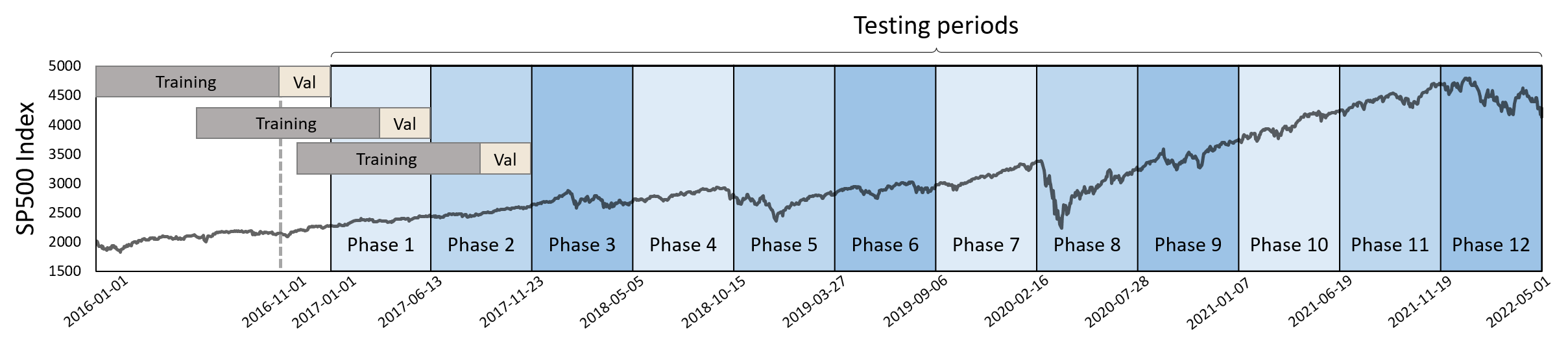}
    \vspace{-2em}
    \caption{Dataset arrangement for backtesting.}
    \label{fig:rolling_backtest}
    \vspace{-1.2em}
\end{figure*}

\sstitle{Wavelet Hypergraph Convolution} To aggregate the extracted temporal 
information of the individual stocks, we develop a hypergraph convolution 
mechanism on the obtained hypergraph $\mathbb{G}$, which consists of multiple 
convolution layers. At each layer $l$, the latent representations of the stocks 
in the previous layer $X^{(l-1)}$ are aggregated by a convolution operator 
HConv(·) using the topology of $\mathbb{G} = (\textbf{\textbf{A}}, 
\textbf{\textbf{W}})$ to generate the current layer representations $X^l$:
\begin{equation}
\mathbf{Z^{(l)}} = \mathrm{HConv}(\mathbf{Z^{(l - 1)}}, \mathbf{A}, \mathbf{W}, \mathbf{P}) 
\label{eqn:conv}
\end{equation}
where $\mathbf{X^l} \in \mathbb{R} ^ {n \times d^l}$ and $\mathbf{X^{l-1}} \in 
\mathbb{R} ^ {n \times d^{l-1}}$ with $n$ being the number of stocks and 
$d^{l-1}, d^l$ as the dimension of the layer-wise latent feature; $\mathbf{P}$ 
is a learnable weight matrix for the layer. 
Following~\cite{yadati2019hypergcn}, the convolution process requires the 
calculation of the hypergraph Laplacian $\mathbf{\Delta}$, which serves as a 
normalized presentation of $\mathbb{G}$:
\begin{equation}
\mathbf{\Delta} = \mathbf{I} - \mathbf{D_v}^{\frac{1}{2}}\mathbf{A}\mathbf{W}\mathbf{D_e}^{-1}\mathbf{A}^T\mathbf{D_v}^{-1/2}
\end{equation}
where $\mathbf{D_v}$ and $\mathbf{D_s}$ are the diagonal matrices containing 
the vertex and hyperedge degrees, respectively. For later usage, we denote 
$\mathbf{D_v}^{\frac{1}{2}}\mathbf{A}\mathbf{W}\mathbf{D_e}^{-1}\mathbf{A}^T\mathbf{D_v}^{-1/2}$
 by $\mathbf{\Theta}$. As $\mathbf{\Delta}$ is a $\mathbb{R}^{n \times n}$ 
positive semi-definite matrix, it can be diagonalized as: $\mathbf{\Delta} = 
\mathbf{U}\mathbf{\Lambda}\mathbf{U^T}$, where $\mathbf{\Lambda} = 
\mathbf{diag}(\lambda_0, \dots, \lambda_k)$ is the diagonal matrix of 
non-negative eigenvalues and $\mathbf{U}$ is the set of orthonormal 
eigenvectors. 

Existing work leverages the Fourier basis~\cite{yadati2019hypergcn} for this 
factorization process. However, using the Fourier basis has two disadvantages: 
(i) the localization during convolution process is not 
well-maintained~\cite{xu2019graph}, and (ii) it requires the direct 
eigen-decomposition of the Laplacian matrix, which is costly for a complex 
hypergraph, such as those faced when modelling stock markets (\textbf{Challenge 
1}). We thus opt to rely on the wavelet basis~\cite{xu2019graph}, for two 
reasons: (i) the wavelet basis represents the information diffusion 
process~\cite{sun2021heterogeneous}, which naturally implements localized 
convolutions of the vertex at each layer, and (ii) the wavelet basis is much 
sparser than the Fourier basis, which enables more efficient computation. 

Applying the wavelet basis, let $\mathbf{\Psi_s} = 
\mathbf{U_s}\mathbf{\Lambda_s}\mathbf{U_s^T}$ be a set of wavelets with scaling 
parameter $-s$. Then, we have $\mathbf{\Lambda_s} = \mathbf{diag}(e^{-\lambda_0 
s}, \dots, e^{-\lambda_k s})$ as the heat kernel matrix. The hypergraph 
convolution process for each vertex $t$ is computed by:
\begin{equation}
HConv(x_t, y) = (\mathbf{\Psi_s}(\mathbf{\Psi_s})^{-1}) \odot (\mathbf{\Psi_s})^{-1} y)= \mathbf{\Psi_s}\mathbf{\Lambda_s}(\mathbf{\Psi_s})^{-1} x_t
\end{equation}
where $y$ is the filter and $(\mathbf{\Psi_s})^{-1} y$ is its corresponding 
spectral transformation. Based on the Stone-Weierstrass 
theorem~\cite{xu2019graph}, the graph wavelet $(\mathbf{\Psi_s})$ can be 
polynomially approximated by:

\begin{equation}
\mathbf{\Psi_s} \approx \sum_{k = 0}^{K}{\alpha_k (\mathbf{\Delta})^k} = \sum_{k = 0}^{K}{\theta_k (\mathbf{\Theta})^k}
\label{eqn:vertex_conv}
\end{equation}
where $K$ is the polynomial order of the approximation. 

The approximation 
facilitates the calculation of $\mathbf{\Psi_s}$ without the 
eigen-decomposition of $\mathbf{\Delta}$. Applying it to 
\autoref{eqn:vertex_conv} and \autoref{eqn:conv} and choosing 
LeakyReLU~\cite{agarap2018deep} as the activation function, we have:
\begin{equation}
\mathbf{Z_H^{(l)}} = LReLU\sum_{k = 0}^K{((\mathbf{D_v}^{\frac{1}{2}}\mathbf{A}\mathbf{W}\mathbf{D_e}^{-1}\mathbf{A}^T\mathbf{D_v}^{-1/2})^k\mathbf{Z_H^{(l-1)}}\mathbf{P})}
\label{eq:final_eq}
\end{equation}

To capture the varying degree of influence each relation between stocks on 
the temporal price evolution of each stock, we also employ an attention 
mechanism~\cite{sawhney2021stock}. This mechanism learns to adaptively weight each 
hyperedge associated with a stock based on its temporal features. For each node 
$v_i \in S$ and its associated hyperedge $e_j \in E$, we compute an attention 
coefficient $\mathbf{\hat{A}}_{ij}$ using the stock’s temporal feature $x_i$ 
and the aggregated hyperedge features $x_j$, quantifying how important the 
corresponding relation $e_j$ is to the stock $v_i$:
\begin{equation}
 \mathbf{\hat{A}}_{ij} = \frac{\mathrm{exp}(LReLU(\hat{a}[\mathbf{P}_{x_i} \mathbin\Vert \mathbf{P}_{x_j}]))}{\sum_{k \in N_i}{\mathrm{exp}(LReLU(\hat{a}[\mathbf{P}_{x_i} \mathbin\Vert \mathbf{P}_{x_j}]))}}
\end{equation}
where $\hat{a}$ is a single-layer feed forward network, $\mathbin\Vert$ is 
concatenation operator and $\mathbf{P}$ represents a learned linear transform. 
$N_i$ is the neighbourhood set of the stock $x_i$, which is derived from the 
constructed hypergraph $\mathbb{G}$. The attention-based learned hypergraph 
incidence matrix $\mathbf{\hat{A}}$ is then used instead of the original 
$\mathbf{A}$ in \autoref{eq:final_eq} to learn intermediate representations of 
the stocks. The representation of the hypergraph is denoted by $\mathbf{Z_H}$, 
which is concatenated with the temporal feature $\mathbf{Z_T}$ to maintain the 
stock individual characteristic (\textbf{Challenge 2}), which then goes through 
the MLP for dimension reduction to obtain the final prediction: 
\begin{equation}
 \mathbf{Z} = MLP(\mathbf{Z_T} \mathbin\Vert \mathbf{Z_H})
\end{equation}
Finally, we use the popular root mean squared error (RMSE) to directly 
encourage the output $\mathbf{X}$ to capture the actual relative price change 
in the short term $d^{(t, t+w)}_s$ of each stock $s$, with $w$ being the 
lookahead window size (with a default value of five).

\section{Empirical Evaluation}
\label{sec:exp}

\begin{table*}[!ht]
\caption{Rolling backtesting from 2017-01-01 to 2022-05-01 on the SP500.} 
\label{tab:rolling_backtest_sp500}
\resizebox{\linewidth}{!}{%
\begin{tabular}{|c|l|c|c|c|c|c|c|c|c|c|c|c|c|c|}
\hline
\multicolumn{1}{|l|}{} & \textbf{Model} & \multicolumn{1}{l|}{\textbf{Phase \#1}} & \multicolumn{1}{l|}{\textbf{Phase \#2}} & \multicolumn{1}{l|}{\textbf{Phase \#3}} & \multicolumn{1}{l|}{\textbf{Phase \#4}} & \multicolumn{1}{l|}{\textbf{Phase \#5}} & \multicolumn{1}{l|}{\textbf{Phase \#6}} & \multicolumn{1}{l|}{\textbf{Phase \#7}} & \multicolumn{1}{l|}{\textbf{Phase \#8}} & \multicolumn{1}{l|}{\textbf{Phase \#9}} & \multicolumn{1}{l|}{\textbf{Phase \#10}} & \multicolumn{1}{l|}{\textbf{Phase \#11}} & \multicolumn{1}{l|}{\textbf{Phase \#12}} & \multicolumn{1}{l|}{\textbf{Mean}} \\ \hline
\multirow{8}{*}{Return} & LSTM & 0.064 & 0.057 & 0.028 & 0.058 & 0.036 & -0.032 & 0.059 & -0.139 & 0.125 & 0.100 & 0.062 & 0.008 & 0.036 \\ \cline{2-15} 
 & ALSTM & 0.056 & 0.043 & 0.022 & 0.053 & 0.009 & -0.068 & 0.036 & -0.121 & 0.115 & 0.097 & 0.066 & 0.009 & 0.026 \\ \cline{2-15} 
 & HATS & 0.102 & 0.031 & -0.003 & 0.042 & 0.062 & 0.067 & 0.092 & -0.074 & 0.188 & 0.132 & 0.063 & -0.059 & 0.054 \\ \cline{2-15} 
 & LSTM-RGCN & 0.089 & 0.051 & 0.005 & -0.006 & 0.077 & 0.019 & 0.088 & -0.121 & 0.155 & 0.107 & 0.038 & -0.032 & 0.039 \\ \cline{2-15} 
 & RSR & 0.065 & 0.043 & -0.009 & -0.016 & 0.014 & -0.007 & 0.059 & -0.113 & 0.089 & 0.056 & 0.038 & -0.052 & 0.014 \\ \cline{2-15} 
 & STHAN-SR & 0.108 & 0.074 & 0.024 & 0.016 & 0.052 & 0.085 & 0.090 & -0.105 & 0.158 & 0.107 & 0.058 & -0.008 & 0.055 \\ \cline{2-15} 
 & HIST & 0.080 & 0.020 & -0.020 & -0.030 & -0.050 & 0.010 & -0.030 & 0.020 & 0.200 & 0.100 & 0.020 & -0.050 & 0.022 \\ \cline{2-15} 
 & \textbf{ESTIMATE} & \textbf{0.109} & \textbf{0.080} & \textbf{0.025} & \textbf{0.105} & \textbf{0.051} & \textbf{0.135} & \textbf{0.149} & \textbf{0.124} & \textbf{0.173} & \textbf{0.065} & \textbf{0.147} & \textbf{0.057} & \textbf{0.102} \\ \hline
\multirow{8}{*}{IC} & LSTM & -0.014 & -0.030 & -0.016 & 0.006 & 0.020 & -0.034 & -0.006 & 0.014 & -0.002 & -0.039 & 0.022 & -0.023 & -0.009 \\ \cline{2-15} 
 & ALSTM & -0.024 & -0.025 & 0.025 & -0.009 & 0.029 & -0.018 & -0.033 & -0.024 & 0.045 & -0.046 & 0.016 & -0.015 & -0.007 \\ \cline{2-15} 
 & HATS & 0.013 & -0.011 & -0.006 & -0.005 & -0.018 & 0.029 & 0.027 & -0.002 & 0.010 & -0.017 & -0.028 & -0.012 & -0.002 \\ \cline{2-15} 
 & LSTM-RGCN & -0.019 & 0.020 & 0.024 & 0.021 & -0.005 & 0.021 & 0.032 & 0.035 & -0.086 & 0.043 & -0.005 & 0.030 & 0.009 \\ \cline{2-15} 
 & RSR & 0.008 & -0.009 & -0.003 & -0.017 & -0.009 & 0.018 & 0.011 & -0.005 & -0.036 & 0.018 & -0.058 & 0.003 & -0.007 \\ \cline{2-15} 
 & STHAN-SR & 0.025 & -0.015 & -0.016 & -0.029 & 0.000 & 0.018 & 0.022 & 0.000 & -0.010 & 0.009 & 0.007 & -0.013 & 0.000 \\ \cline{2-15} 
 & HIST & 0.003 & 0.000 & 0.005 & -0.010 & 0.006 & 0.008 & 0.005 & -0.017 & 0.006 & 0.009 & 0.011 & 0.006 & 0.003 \\ \cline{2-15} 
 & \textbf{ESTIMATE} & \textbf{0.037} & \textbf{0.080} & \textbf{0.153} & \textbf{0.010} & \textbf{0.076} & \textbf{0.080} & \textbf{0.080} & \textbf{0.011} & \textbf{0.127} & \textbf{0.166} & \textbf{0.010} & \textbf{0.131} & \textbf{0.080} \\ \hline
\multirow{8}{*}{Rank\_IC} & LSTM & -0.151 & -0.356 & -0.289 & 0.089 & 0.186 & -1.091 & -0.151 & 0.201 & -0.019 & -0.496 & 0.259 & -0.397 & -0.185 \\ \cline{2-15} 
 & ALSTM & -0.211 & -0.266 & 0.409 & -0.099 & 0.182 & -0.289 & -0.476 & -0.243 & 0.242 & -0.323 & 0.094 & -0.174 & -0.096 \\ \cline{2-15} 
 & HATS & 0.169 & -0.156 & -0.139 & -0.063 & -0.408 & 0.517 & 0.333 & -0.032 & 0.085 & -0.344 & -0.547 & -0.135 & -0.060 \\ \cline{2-15} 
 & LSTM-RGCN & -0.271 & 0.210 & 0.223 & 0.152 & -0.035 & 0.279 & 0.261 & 0.273 & -0.416 & 0.329 & -0.036 & 0.354 & 0.110 \\ \cline{2-15} 
 & RSR & 0.151 & -0.159 & -0.051 & -0.213 & -0.107 & 0.282 & 0.135 & -0.090 & -0.292 & 0.175 & -0.541 & 0.040 & -0.056 \\ \cline{2-15} 
 & STHAN-SR & 0.690 & -0.357 & -0.365 & -0.714 & 0.008 & 0.369 & 0.523 & 0.005 & -0.265 & 0.169 & 0.141 & -0.276 & -0.006 \\ \cline{2-15} 
 & HIST & 0.085 & -0.008 & 0.125 & -0.225 & 0.192 & 0.204 & 0.107 & -0.328 & 0.174 & 0.256 & 0.215 & 0.157 & 0.080 \\ \cline{2-15} 
 & \textbf{ESTIMATE} & \textbf{0.386} & \textbf{0.507} & \textbf{1.613} & \textbf{0.059} & \textbf{0.284} & \textbf{0.585} & \textbf{0.412} & \textbf{0.062} & \textbf{0.704} & \textbf{0.936} & \textbf{0.054} & \textbf{0.595} & \textbf{0.516} \\ \hline
\multirow{8}{*}{ICIR} & LSTM & -0.010 & -0.036 & -0.007 & 0.010 & 0.016 & -0.038 & 0.004 & 0.010 & 0.011 & -0.041 & 0.021 & -0.010 & -0.006 \\ \cline{2-15} 
 & ALSTM & -0.057 & -0.041 & 0.030 & -0.012 & 0.033 & -0.015 & -0.028 & -0.034 & 0.057 & -0.053 & 0.009 & -0.002 & -0.009 \\ \cline{2-15} 
 & HATS & 0.023 & -0.014 & 0.010 & 0.001 & -0.016 & 0.033 & 0.035 & 0.020 & 0.023 & -0.005 & -0.042 & -0.034 & 0.003 \\ \cline{2-15} 
 & LSTM-RGCN & -0.033 & 0.012 & 0.022 & 0.027 & -0.009 & 0.028 & 0.047 & 0.051 & -0.085 & 0.054 & -0.006 & 0.039 & 0.012 \\ \cline{2-15} 
 & RSR & 0.031 & -0.018 & -0.005 & -0.033 & -0.009 & 0.029 & 0.001 & -0.007 & -0 .019 & 0.017 & -0.072 & -0.031 & -0.010 \\ \cline{2-15} 
 & STHAN-SR & 0.018 & -0.011 & -0.016 & -0.021 & 0.005 & 0.016 & 0.023 & 0.008 & -0.003 & 0.004 & 0.009 & -0.007 & 0.002 \\ \cline{2-15} 
 & HIST & 0.004 & -0.002 & -0.006 & -0.001 & 0.007 & 0.001 & \textbf{0.005} & -0.014 & 0.009 & 0.009 & 0.021 & 0.010 & 0.004 \\ \cline{2-15} 
 & \textbf{ESTIMATE} & \textbf{0.033} & \textbf{0.081} & \textbf{0.148} & \textbf{0.032} & \textbf{0.076} & \textbf{0.103} & \textbf{0.064} & \textbf{0.058} & \textbf{0.103} & \textbf{0.142} & \textbf{0.020} & \textbf{0.098} & \textbf{0.080} \\ \hline
\multirow{8}{*}{Rank\_ICIR} & LSTM & -0.100 & -0.364 & -0.110 & 0.140 & 0.141 & -0.984 & 0.094 & 0.168 & 0.117 & -0.525 & 0.227 & -0.172 & -0.114 \\ \cline{2-15} 
 & ALSTM & -0.423 & -0.344 & 0.415 & -0.134 & 0.202 & -0.192 & -0.313 & -0.343 & 0.306 & -0.377 & 0.047 & -0.019 & -0.098 \\ \cline{2-15} 
 & HATS & 0.234 & -0.155 & 0.179 & 0.013 & -0.221 & 0.511 & 0.340 & 0.270 & 0.194 & -0.076 & -0.525 & -0.372 & 0.033 \\ \cline{2-15} 
 & LSTM-RGCN & -0.387 & 0.111 & 0.197 & 0.170 & -0.058 & 0.313 & 0.284 & 0.326 & -0.354 & 0.318 & -0.034 & 0.427 & 0.109 \\ \cline{2-15} 
 & RSR & 0.378 & -0.353 & -0.053 & -0.285 & -0.065 & 0.326 & 0.010 & -0.081 & -0.114 & 0.131 & -0.428 & -0.288 & -0.068 \\ \cline{2-15} 
 & STHAN-SR & 0.435 & -0.211 & -0.310 & -0.573 & 0.107 & 0.386 & 0.478 & 0.271 & -0.068 & 0.075 & 0.230 & -0.149 & 0.056 \\ \cline{2-15} 
 & HIST & 0.079 & -0.044 & -0.144 & -0.020 & 0.205 & 0.018 & 0.122 & -0.289 & 0.236 & 0.229 & 0.356 & 0.209 & 0.080 \\ \cline{2-15} 
 & \textbf{ESTIMATE} & \textbf{0.315} & \textbf{0.446} & \textbf{1.344} & \textbf{0.178} & \textbf{0.307} & \textbf{0.587} & \textbf{0.329} & \textbf{0.311} & \textbf{0.541} & \textbf{0.885} & \textbf{0.100} & \textbf{0.488} & \textbf{0.486} \\ \hline
\multirow{8}{*}{Prec@N} & LSTM & 0.542 & 0.553 & 0.581 & 0.471 & 0.456 & 0.569 & 0.440 & 0.547 & 0.588 & 0.615 & 0.554 & 0.608 & 0.544 \\ \cline{2-15} 
 & ALSTM & 0.583 & 0.585 & 0.550 & 0.471 & 0.514 & 0.575 & 0.431 & 0.556 & 0.650 & 0.518 & 0.497 & 0.627 & 0.546 \\ \cline{2-15} 
 & HATS & 0.624 & 0.651 & 0.597 & 0.495 & 0.551 & 0.642 & 0.532 & 0.619 & 0.542 & 0.550 & 0.529 & 0.690 & 0.585 \\ \cline{2-15} 
 & LSTM-RGCN & 0.565 & 0.589 & 0.600 & 0.505 & 0.505 & 0.628 & 0.522 & 0.583 & 0.517 & 0.538 & 0.566 & 0.592 & 0.559 \\ \cline{2-15} 
 & RSR & 0.587 & 0.531 & 0.608 & 0.455 & 0.465 & 0.619 & 0.473 & 0.608 & 0.553 & 0.618 & 0.590 & 0.676 & 0.565 \\ \cline{2-15} 
 & STHAN-SR & 0.554 & 0.614 & 0.573 & 0.463 & 0.562 & 0.611 & 0.530 & 0.575 & 0.553 & 0.626 & 0.534 & 0.510 & 0.559 \\ \cline{2-15} 
 & HIST & 0.691 & 0.625 & 0.476 & 0.512 & 0.452 & 0.561 & 0.548 & 0.634 & 0.463 & 0.615 & 0.395 & 0.605 & 0.548 \\ \cline{2-15} 
 & \textbf{ESTIMATE} & \textbf{0.619} & \textbf{0.631} & \textbf{0.673} & \textbf{0.524} & \textbf{0.540} & \textbf{0.739} & \textbf{0.568} & \textbf{0.669} & \textbf{0.611} & \textbf{0.679} & \textbf{0.547} & \textbf{0.724} & \textbf{0.627} \\ \hline
\end{tabular}%
}    
\end{table*}

In this section, we empirically evaluate our framework based on four research questions, as follows:
\begin{compactitem}
\item[(RQ1)] Does our model outperform the baseline methods?
\item[(RQ2)] What is the influence of
each model component?
\item[(RQ3)] Can our model be interpreted in a qualitative sense? 
\item[(RQ4)] Is our model sensitive to hyperparameters?
\end{compactitem}
Below, we first describe the experimental setting (\autoref{sec:exp_setup}). We 
then present our empirical evaluations, including an end-to-end comparison 
(\autoref{sec:exp_accuracy}), a qualitative study 
(\autoref{sec:exp_qualitative}), an ablation test (\autoref{sec:exp_ablation}), 
and an examination of the hyperparameter sensitivity 
(\autoref{sec:exp_sensitivity}).

\subsection{Setting}
\label{sec:exp_setup}

\sstitle{Datasets} We evaluate our approach based on the US stock market. We 
gathered historic price data and the information about industries in the S\&P 
500 index from the Yahoo Finance database 
\cite{yahoo-finance}, covering 2016/01/01 to 2022/05/01 (1593 
trading days). Overall, while the market witnessed an upward trend in this 
period, it also experienced some considerable correction in 2018, 2020, and 
2022. We split the data of this period into 12 phases with varying degrees of 
volatility, with the period between two consecutive phases being 163 days. Each 
phase contains 10 month of training data, 2 month of validation data, and 6 
month of testing data (see \autoref{fig:rolling_backtest}).

\sstitle{Metrics} We adopt the following evaluation metrics:
To thoroughly evaluate the performance of the techniques, we employ the 
following metrics:
\begin{compactitem}
    \item \emph{Return:} is the estimated profit/loss ratio that the portfolio 
    achieves after a specific period, calculated by $NV_e / NV_s - 1$, with 
    $NV_s$ and $NV_e$ being the net asset value of the portfolio before and 
    after the period. 

    \item \emph{Information Coefficient (IC):} is a coefficient that shows how 
    close the prediction is to the actual result, computed by the average 
    Pearson correlation coefficient. 
    
    \item \emph{Information ratio based IC (ICIR):} The information ratio of the IC metric, calculated by $ICIR = mean(IC)/std(IC)$

    \item \emph{Rank Information Coefficient (Rank\_IC):} is the 
    coefficient based on the ranking of the stocks' short-term profit 
    potential, computed by the average Spearman 
    coefficient~\cite{myers2004spearman}.

    \item \emph{Rank\_ICIR:} {Information ratio based Rank\_IC (ICIR):} The information ratio of the Rank\_IC metric, calculated by: \\ $rank\_ICIR = mean(rank\_IC)/std(rank\_IC)$
    
    \item \emph{Prec@N:} evaluates the precision of the top N short-term profit 
    predictions from the model. This way, we assess 
    the capability of the techniques to support investment decisions.
\end{compactitem}

\sstitle{Baselines} 
We compared the performance of our technique with that of several state-of-the-art baselines, as follows:
\begin{compactitem}
    \item \emph{LSTM:} \cite{lstm} is a traditional baseline which leverages a 
    vanilla LSTM on temporal price data. 
    \item \emph{ALSTM:} \cite{alstm} is a stock movement prediction framework 
    that integrates the adversarial training and stochasticity simulation in an 
    LSTM to better learn the market dynamics.

    \item \emph{HATS:} \cite{kim2019hats} is a stock prediction framework that 
    models the market as a classic heterogeneous graph and propose a 
    hierarchical graph attention network to learn a stock representation to 
    classify next-day movements.
    
    
    \item \emph{LSTM-RGCN:} \cite{li2020modeling} is a graph-based prediction framework that constructs the connection among stocks with their price 
    correlation matrix and learns the spatio-temporal relations using a 
    GCN-based encoder-decoder architecture. 
    
    \item \emph{RSR:} \cite{rsr} is a stock prediction framework that combines 
    Temporal Graph Convolution with LSTM to learn the stocks' relations in a 
    time-sensitive manner. 
    
    \item \emph{HIST:} \cite{xu2021hist} is a graph-based stock trend forecasting framework that follows the encoder-decoder paradigm in attempt to capture the shared information between stocks from both predefined concepts as well as revealing hidden concepts. 
    
    \item \emph{STHAN-SR:} \cite{sawhney2021stock} is a deep learning-based framework that also models the complex relation of the stock market as a hypergraph and employs vanilla hypergraph convolution to learn directly the stock short-term profit ranking.

\end{compactitem}    

\sstitle{Trading simulation} We simulate a trading portfolio using the output 
prediction of the techniques. 
At each timestep, the portfolio allocates an equal portion of money for k 
stocks, as determined by the prediction. 
We simulate the risk control by applying a trailing stop level of 7\% and 
profit 
taking level of 20\% for all positions. 
We ran the simulation 1000 times per phase and report average 
results.   

\sstitle{Reproducibility environment} 
All experiments were conducted on an AMD Ryzen ThreadRipper 3.8 GHz system with 
128 GB of main memory and four RTX 3080 graphic cards. We used Pytorch for the 
implementation and Adam as gradient optimizer.

\subsection{End-to-end comparisons}
\label{sec:exp_accuracy}

\begin{figure*}[!h]
	\centering
	\vspace{-1em}
	\includegraphics[width=.7\linewidth]{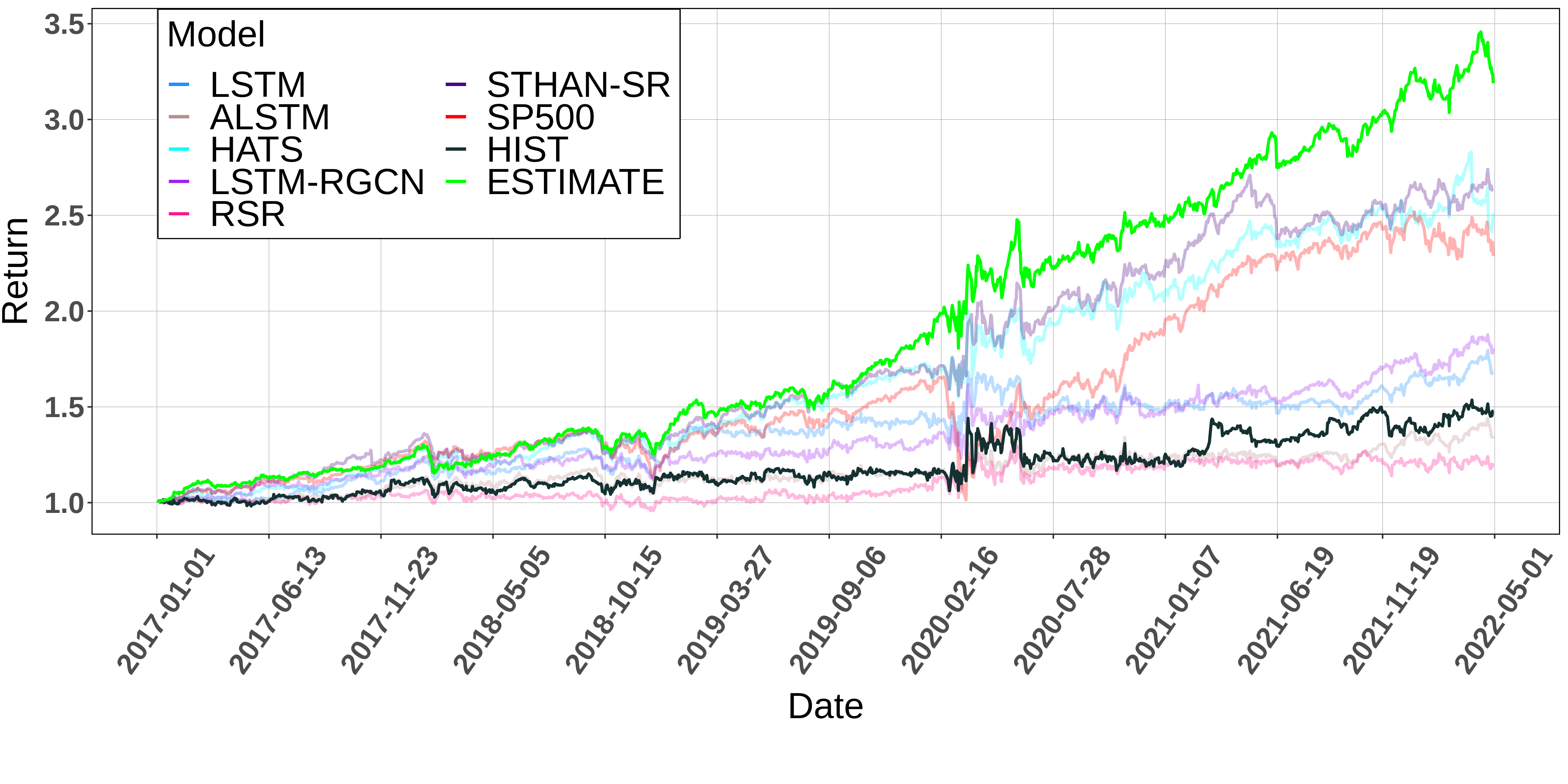}
	\vspace{-2em}
	\caption{Cumulative return.}
	\label{fig:cum_return}
	\vspace{-1em}
\end{figure*}

To answer research question RQ1, we report in 
\autoref{tab:rolling_backtest_sp500} an 
end-to-end comparison of our approach (ESTIMATE) against the baseline methods. 
We also visualize the average accumulated return of the baselines and the S\&P 
500 index during all 12 phases in \autoref{fig:cum_return}.

In general, our model outperforms all baseline methods across all datasets in terms of \emph{Return}, \emph{IC}, \emph{Rank\_IC} and \emph{Prec@10}. Our technique consistently 
achieves a positive return and an average \emph{Prec@10} of 0.627 over all 12 
phases; 
and performs significant better than the S\&P 500 index with higher overall 
return. STHAN-SR is the best method 
among the baselines, yielding a high \emph{Return} in some phases (\#1, 
\#2, \#5, \#10). This is because STHAN-SR, similar to our approach, models the 
multi-order relations 
between the stocks using a hypergraph. However, our technique still outperforms 
STHAN-SR by a considerable margin for the other periods, including the ranking 
metric \emph{Rank\_IC} even though our technique does not aim to learn directly the 
stock rank, like STHAN-SR. 

Among the other baseline, models with a relational basis like RSR, 
HATS, and LSTM-RGCN outperform vanilla LSTM and ALSTM models. However, the gap 
between classic graph-based techniques like HATS and LSTM-RGCN is small. This 
indicates that the complex relations in a stock market shall be modelled. An 
interesting finding is that all the performance of the 
techniques drops significantly during phase \#3 and phase \#4, even though the 
market moves sideways. This observation highlights issues of 
prediction algorithm when there is no clear trend for the market.

\begin{figure}[!h]
	\centering
	\vspace{-1em}
	\includegraphics[width=1\linewidth]{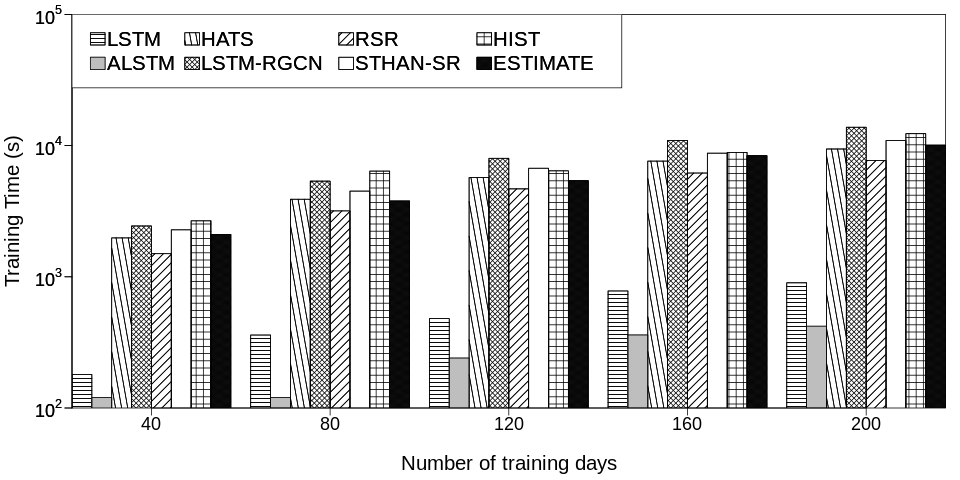}
	\vspace{-2em}
	\caption{Training performance.}
	\label{fig:time}
	\vspace{-1em}
\end{figure}

The training times for these techniques are shown in \autoref{fig:time}, where 
we consider the size of training set ranging from 40 to 200 days. As expected, 
the graph-based techniques (HATS, LSTM-RGCN, RSR, STHAN-R, HIST and ESTIMATE) are 
slower than the rest, due to the trade-off between accuracy and computation 
time. Among the graph-based techniques, there is no significant difference 
between the techniques using classic graphs (HATS, LSTM-RGCN) and those using 
hypergraphs (ESTIMATE, STHAN-R). Our technique ESTIMATE is faster than STHAN-R 
by a considerable margin and is one of the fastest among the graph-based 
baselines, which highlights the efficiency of our wavelet convolution scheme 
compared to the traditional Fourier basis.

\subsection{Ablation Study} 
\label{sec:exp_ablation}

To answer question RQ2, we evaluated the importance of individual components of 
our 
model by creating four variants:
\emph{(EST-1)} This variant does not employ the hypergraph 
convolution, but directly uses the extracted temporal features to predict the 
short-term trend of stock prices.
\emph{(EST-2)} This variant does not employ the generative filters,  
but relies on a common attention LSTM by existing approaches.
\emph{(EST-3)} This variant does not apply the price-correlation 
based augmentation, as described in \autoref{sec:market_agg}. It employs solely 
the industry-based hypergraph as input. 
\emph{(EST-4)} This variant does not employ the wavelet basis for 
hypergraph convolution, as introduced in \autoref{sec:market_agg}. Rather, the 
traditional Fourier basis is applied.

\begin{table}[!h]
\centering
\footnotesize
\vspace{-1em}
\caption{Ablation test}
\label{tbl:ablation}
\vspace{-1em}
\scalebox{1}{
\begin{tabular}{@{} l l l l l l @{}}
\toprule
\textbf{Metric} & \textbf{ESTIMATE} & \textbf{EST-1} & \textbf{EST-2} & 
\textbf{EST-3} & \textbf{EST-4}  \\
\midrule
Return & \textbf{0.102} & 0.024 & 0.043 & 0.047 & 0.052 \\
IC & \textbf{0.080} & 0.013 & 0.020 & 0.033 & 0.020 \\
RankIC & \textbf{0.516} & 0.121 & 0.152 & 0.339 & 0.199 \\
Prec@N & \textbf{0.627} & 0.526 & 0.583 & 0.603 & 0.556 \\
\bottomrule
\end{tabular}
}
\vspace{-1em}
\end{table}

\autoref{tbl:ablation} presents the results for several evaluation metrics, 
averaged over all phases due to space constraints. We observe that our full 
model 
\emph{ESTIMATE} outperforms the other variants, which provides evidence for the 
the positive impact of each of its components. In particular, it is 
unsurprising that the removal of the relations between stocks leads to a 
significant degradation of the final result (approximately 
75\% of the average return) in \textit{EST-1}. A similar drop of the average 
return can 
be seen for \emph{EST-2} and \emph{EST-3}, which highlights the benefits of 
using generati
+ve filters over a traditional single LSTM temporal extractor 
(\emph{EST-2}); and of the proper construction of the representative hypergraph 
(\emph{EST-3}). Also, the full model outperforms the variant \emph{EST-4} by a 
large margin in every metric. This underlines the robustness of the 
convolution with the wavelet basis used in \emph{ESTIMATE} over the traditional 
Fourier basis that is used in existing work.

\subsection{Qualitative Study}

\label{sec:exp_qualitative}


\begin{figure*}[!h]
\centering
	\begin{subfigure}[b]{0.4\textwidth}
         \centering
         \includegraphics[width=1.0\linewidth]{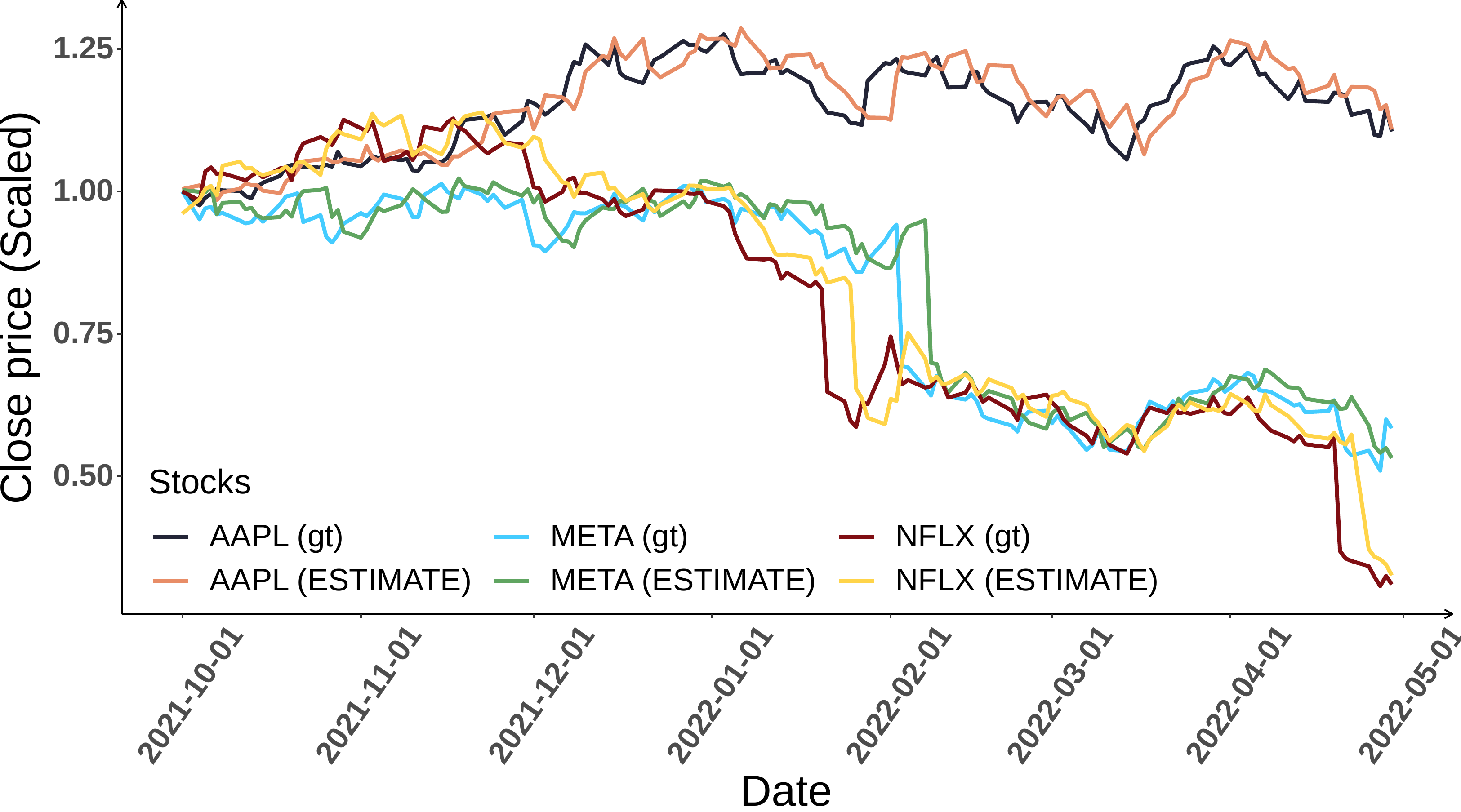}
         \caption{ESTIMATE}
         \label{fig:trend_0}
     \end{subfigure}
     \begin{subfigure}[b]{0.4\textwidth}
         \centering
         \includegraphics[width=1.0\linewidth]{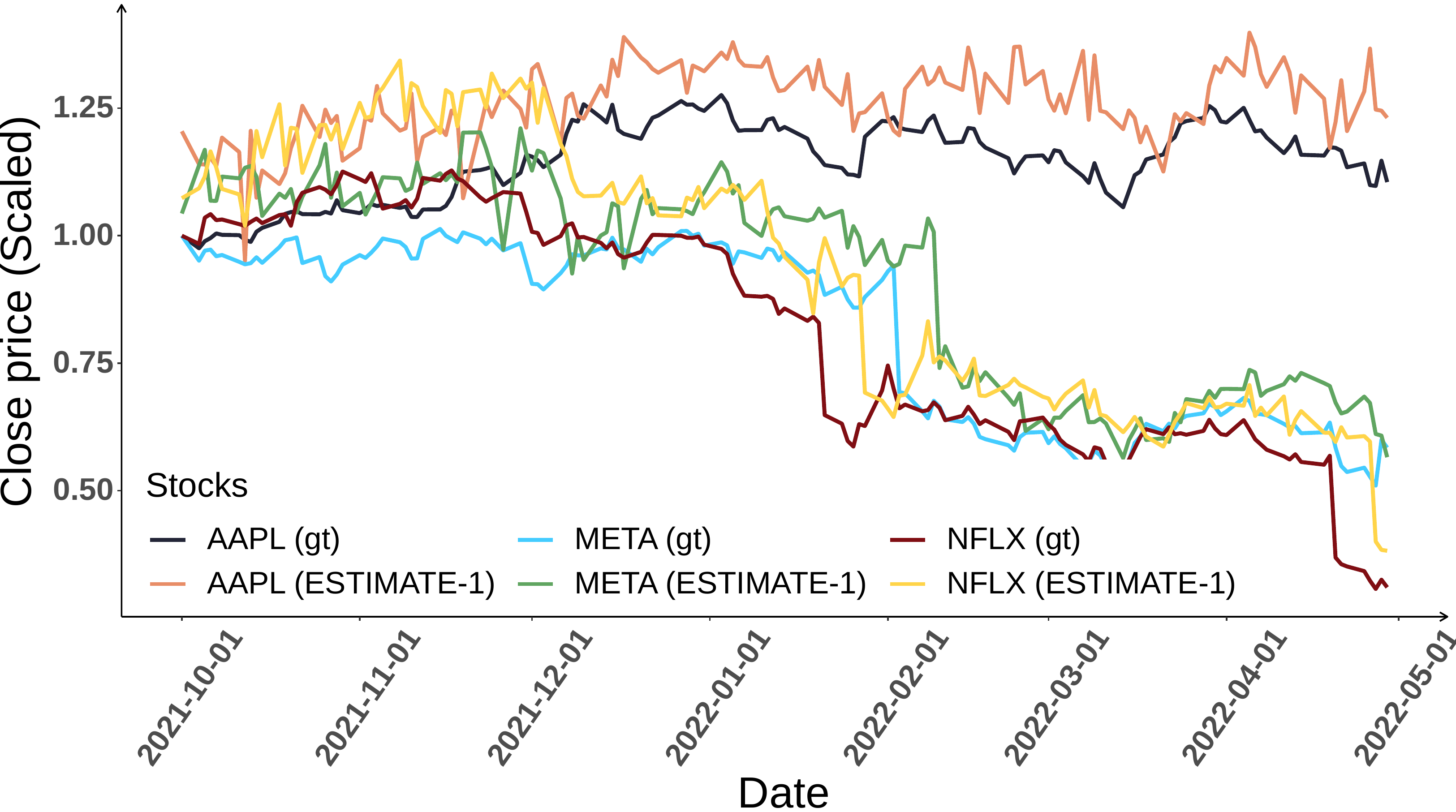}
         \caption{ESTIMATE-1}
         \label{fig:trend_1}
     \end{subfigure}
     \begin{subfigure}[b]{0.4\textwidth}
         \centering
         \includegraphics[width=1.0\linewidth]{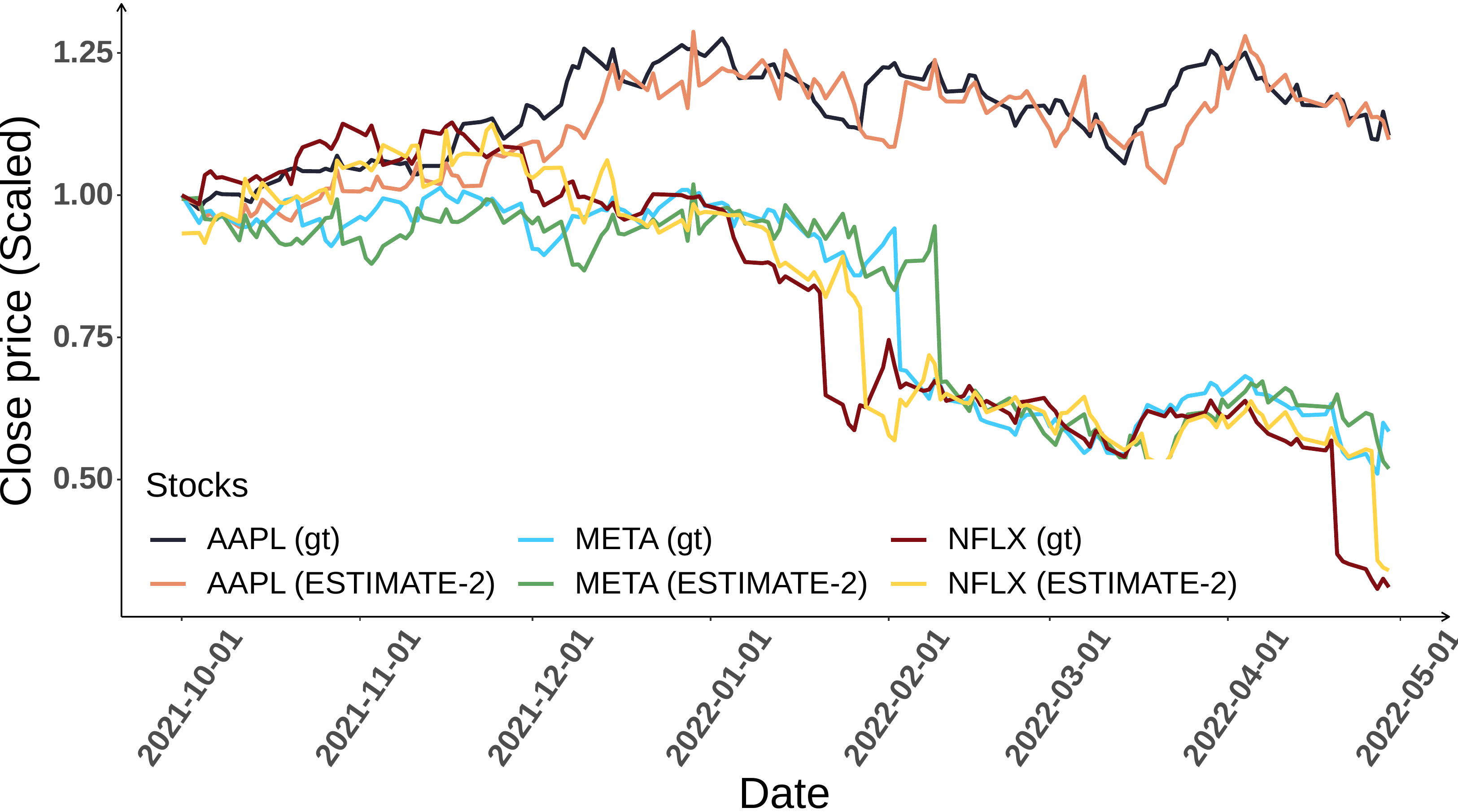}
         \caption{ESTIMATE-2}
         \label{fig:trend_2}
     \end{subfigure}
	\vspace{-1em}
\caption{Trend prediction}
\label{fig:trend_prediction}
\vspace{-1em}
\end{figure*}

We answer research question RQ3 by visualizing in 
\autoref{fig:trend_prediction} the prediction 
results of our technique ESTIMATE for the technology stocks APPL and META from 
01/10/2021 to 01/05/2022. We also compare ESTIMATE's performance to the variant
that does not consider relations between stocks (EST-1) and the variant that 
does not employ temporal generative filters (EST-2). This way, we illustrate 
how our technique is able to handle \textbf{Challenge 1} and \textbf{Challenge 
2}. 

The results indicate that modelling the complex multi-order dynamics of a stock 
market (\textbf{Challenge 
1}) helps ESTIMATE and EST-2 to correctly predict the downward trend of 
technology stocks around the start of 2022; while the prediction of EST-1, 
which uses the temporal patterns of each stock, suffers from a 
significant delay. Also, the awareness of internal dynamics of ESTIMATE due to 
the usage of generative filters helps our technique to differentiate the trend 
observed for APPL from the one of META, especially at the start of the 
correction period in January 2022. 

\subsection{Hyperparameter sensitivity}
\label{sec:exp_sensitivity}

This experiment addresses question RQ4 on the hyperparameter sensitivity. Due 
to space limitations, we focus on the most important hyperparameters. The 
backtesting period of this experiment is set from 01/07/2021 to 01/05/2022 for 
the same reason. 

\begin{figure*}[!h]
\centering
	\begin{subfigure}[b]{0.4\textwidth}
         \centering
         \includegraphics[width=1.0\linewidth]{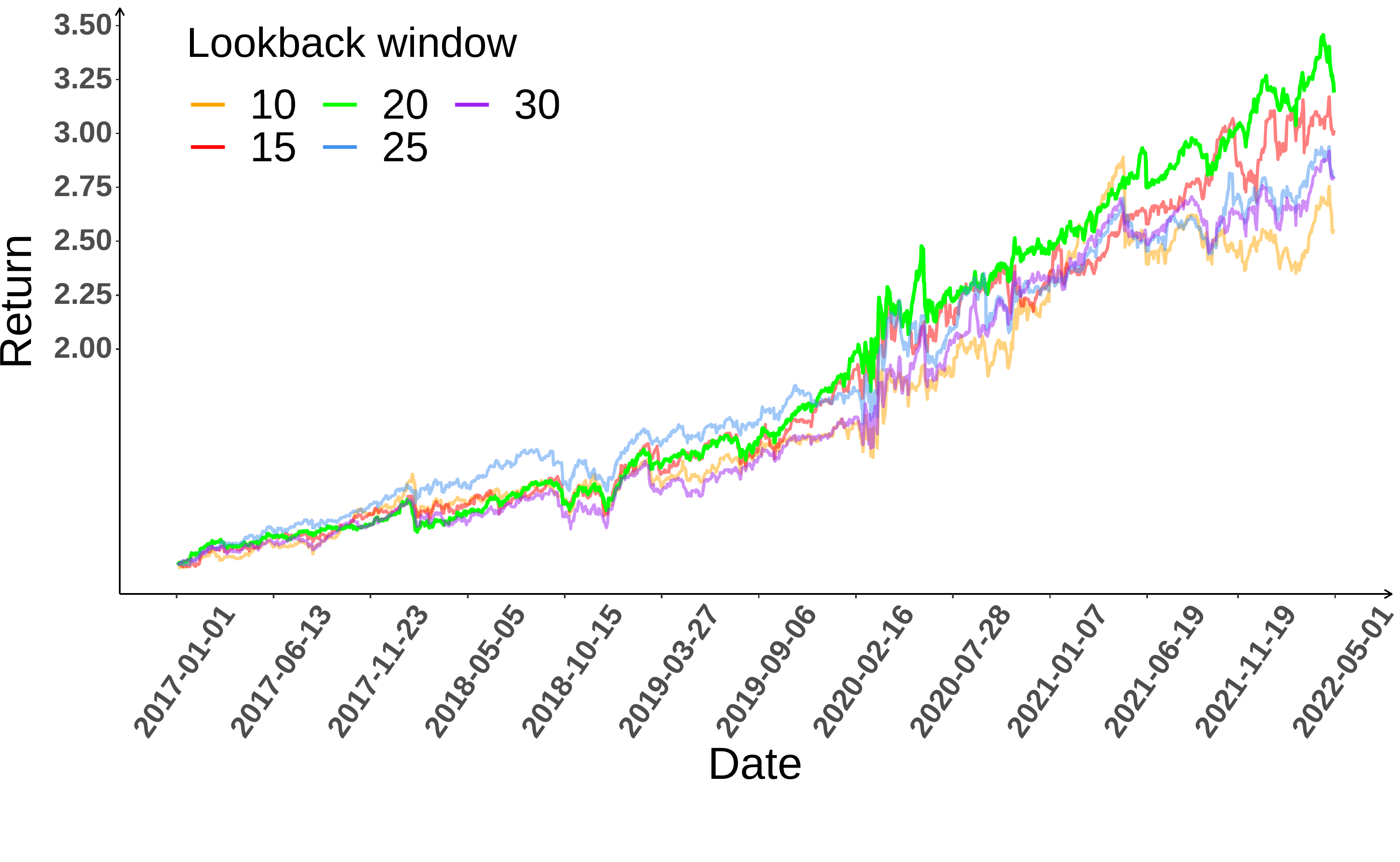}
         \vspace{-1em}
         \caption{Lookback window}
         \label{fig:hyper_lbw}
     \end{subfigure}
     \begin{subfigure}[b]{0.4\textwidth}
         \centering
         \includegraphics[width=1.0\linewidth]{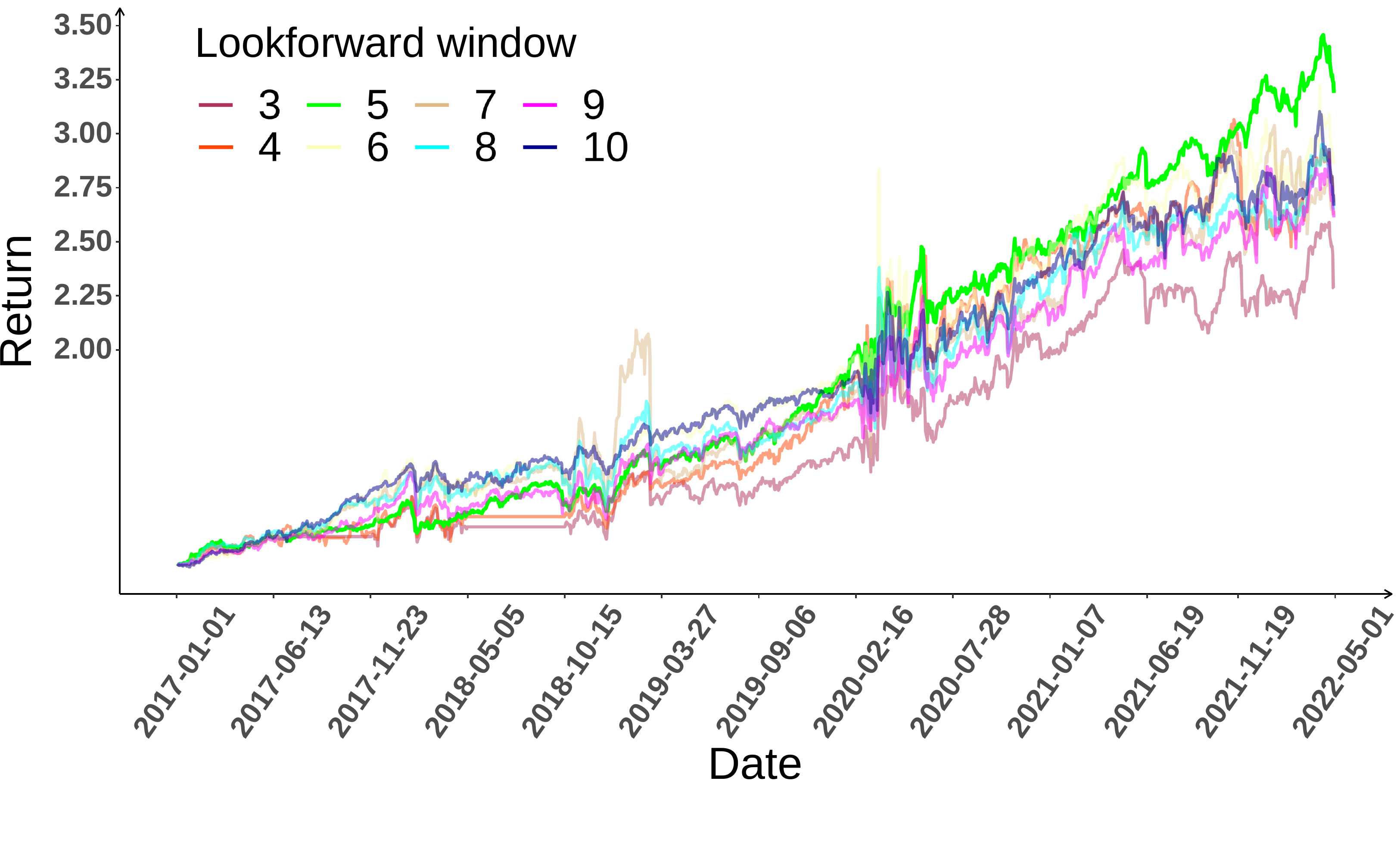}
         \caption{Lookforward window}
         \label{fig:hyper_lfw}
     \end{subfigure}
     \begin{subfigure}[b]{0.4\textwidth}
         \centering
         \includegraphics[width=1.0\linewidth]{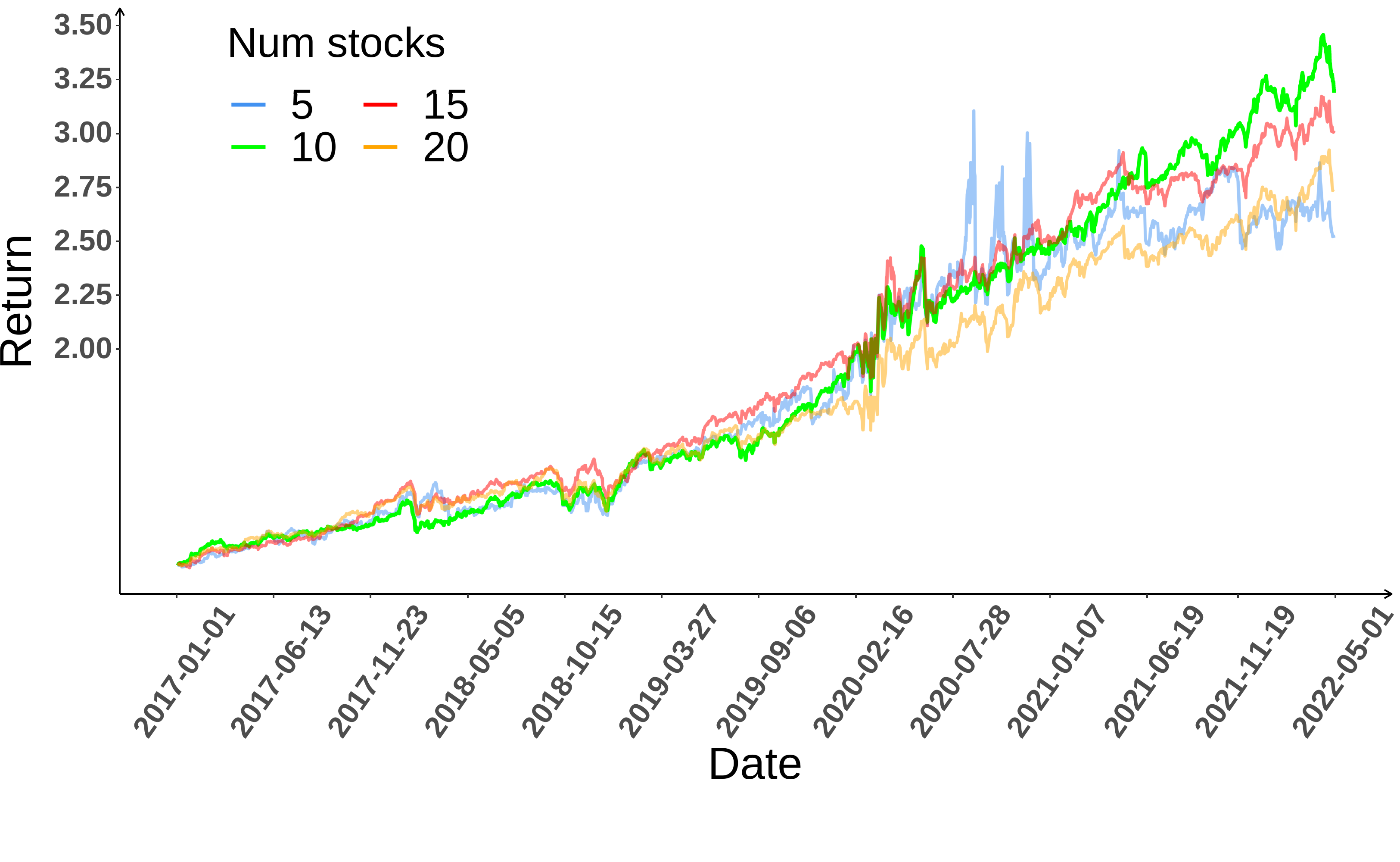}
         \caption{Num stocks}
         \label{fig:hyper_ns}
     \end{subfigure}
	\vspace{-1em}
\caption{Hyperparameters sensitivity}
\label{fig:hyper}
	\vspace{-1em}
\end{figure*}

\sstitle{Lookback window length T} We analyze the prediction performance of 
ESTIMATE when varying the length T of the lookback window in 
\autoref{fig:hyper}. It can be observed that the best window length is 20, 
which coincides with an important length commonly used by professional analyses 
strategies~\cite{adam2016stock}. The performance drops quickly when the window 
length is less or equal than 10, due to the lack of information. On the other 
hand, the performance also degrades when the window length increases above 25. 
This shows that even when using an LSTM to mitigate the vanishing gradient 
issue, the model cannot handle very long sequences. 

\sstitle{Lookahead window length w} We consider different lengths w of the 
lookahead window (\autoref{fig:hyper}) and observe that ESTIMATE achieves the 
best performance for a window of length 5. The results degrade significantly 
when w exceeds 10. This shows that our model performs well for short-term 
prediction, it faces issues when considering a long-term view. 

\sstitle{Number of selected top-k stocks} We analyze the variance in the 
profitability depending on the number of selected top-k stocks from the ranking 
in \autoref{fig:hyper}. We find that ESTIMATE performs generally well, while 
the best results are obtained for k = 10.


\section{Related Work}
\label{sec:related}

\sstitle{Traditional Stock Modelling} Traditional techniques often focus on 
numerical features~\cite{liou2021finsense, ruiz2012correlating}, referred to as 
technical indicators, such as the Moving Average (MA) or the Relative Strength 
Index (RSI). The features are combined with classical timeseries models, such 
as ARIMA~\cite{piccolo1990distance}, to model the stock 
movement~\cite{ariyo2014stock}. However, such techniques often require the 
careful engineering by experts to identify effective indicator combinations and 
thresholds. Yet, these configurations are often not robust against market 
changes. 

For individual traders, they often engineer rules based on a specific set of technical indicators (which indicate the trading momentum) to find the buying signals. For instance, a popular strategy is to buy when the moving average of length 5 (MA5) cross above the moving average of length 20 (MA20), and the Moving Average Convergence Divergence (MACD) is positive. However, the engineering of the specific features requires the extensive expertise and experiment of the traders in the market. Also, the traders must continuously tune and backtest the strategy, as the optimized strategy is not only different from the markets but also keeps changing to the evolving of the market. Last but not least, it is exhaustive for the trader to keep track of a large number of indicators of a stock, as well as the movement of multiple stocks in real time. Due to these drawbacks, quantitative trading with the aid of the AI emerges in recent years, especially with the advances of deep neural network (DNN).

\sstitle{DNN-based Stock Modelling} Recent techniques leverage advances 
in deep learning to capture the non-linear temporal dynamics of 
stock prices through high-level latent 
features~\cite{ren2019investment,huynh2021network,bendre2021gpr}. Earlier 
techniques following 
this paradigm employ recurrent neural networks (RNN)~\cite{ nelson2017stock} or 
convolutional neural networks (CNN)~\cite{tsantekidis2017forecasting} to model 
a single stock price and predict its short-term trend. Other works employ 
deep reinforcement learning (DRL), the combination of deep learning with 
reinforcement learning (RL), a subfield of sequential decision-making. For 
instance, the quantitative trading 
problem can be formulated as a Markov decision process~\cite{LiuYGW21} and 
be addressed by well-known DRL algorithms (e.g. DQN, 
DDPG~\cite{lillicrap2015continuous}). However, these techniques treat the 
stocks independently and lack a proper scheme to consider the 
complex relations between stocks in the market. 

\sstitle{Graph-based Stock Modelling} Some state-of-the-art techniques address 
the problem of correlations between stocks and propose graph-based solutions to 
capture the inter-stock 
relations. For instance, the market may be modelled as a heterogeneous 
graph with different type of pairwise relations~\cite{kim2019hats}, which is 
then used in an 
attention-based graph convolution network (GCN)~\cite{tam2023pr,nguyen2022model,nguyen2021structural,nguyen2020entity,tam2021entity,trung2020adaptive} to predict the stock 
price and market index movement. Similarly, a 
market graph may be constructed and an augmented GCN with temporal convolutions 
can be employed to learn, at the same 
time, the stock movement and stock relation evolution~\cite{li2020modeling}. 
The most recent techniques~\cite{rsr,sawhney2021stock} are based on the argument that the 
stock market includes multi-order relations, so that the market should be 
modelled using hypergraphs. Specifically, external knowledge from knowledge 
graphs enables the construction of a  
market hypergraph~\cite{sawhney2021stock}, which is used in a spatiotemporal 
attention hypergraph network to learn interdependences of stocks and their 
evolution. Then, a ranking of stocks based on short-term 
profit is derived. 

Different from previous work, we propose a market analysis framework that 
learns the complex multi-order correlation of stocks derived from a 
hypergraph representation. We go beyond the state of the art (\cite{rsr,sawhney2021stock}) by proposing temporal generative filters that implement a memory-based 
mechanism to recognize better the individual characteristics of each stock,  
while not over-parameterizing the core LSTM model. Also, we propose a new 
hypergraph attention convolution scheme that leverages the wavelet basis to 
mitigate the high complexity and dispersed localization faced in previous 
hypergraph-based approaches.

\section{Conclusion}
\label{sec:con}

In this paper, we address two unique characteristics of the stock market 
prediction problem: (i) \emph{multi-order dynamics} which implies strong 
non-pairwise correlations between the price movement of different stocks, and 
(ii) \emph{internal dynamics} where each stock maintains its own dedicated 
behaviour. We propose \textit{ESTIMATE}, a stock recommendation framework that 
supports learning of the multi-order correlation of the stocks (i) and their 
individual temporal patterns (ii), which are then encoded in node embeddings 
derived from hypergraph representations. 
The framework provides two novel 
mechanisms: First, \textit{temporal generative filters} are incorporated as a 
memory-based shared parameter LSTM network that facilitates learning of 
temporal patterns per stock. Second, we presented attention hypergraph 
convolutional layers using the wavelet basis, i.e., a convolution paradigm that 
relies on the polynomial wavelet basis to simplify the message passing and 
focus on the localized convolution. 

Extensive experiments on real-world data illustrate the effectiveness of our 
techniques and highlight its 
applicability in trading recommendation. Yet, the experiments also illustrate 
the impact of concept drift, when the market characteristics change from the 
training to the testing period. 
In future work, we plan to tackle this issue by exploring 
time-evolving hypergraphs with the ability to memorize distinct periods of 
past data and by incorporating external data sources such as earning calls, 
fundamental indicators, news data~\cite{nguyen2019user,nguyen2019maximal,nguyen2015result}, social networks~\cite{nguyen2022detecting,tam2019anomaly,nguyen2017retaining,nguyen2021judo}, and crowd signals~\cite{nguyen2013batc,hung2017computing,hung2013evaluation}.

%



%
%





\appendix
\section{Technical indicators formulation}
In this section, we express equations to formulate indicators described in \textbf{Section 3: Temporal Generative Filters.}
Let $t$ denote the $t$-th time step and $O_t, H_t, L_t, C_t, V_t$ represents the open price, high price, low price, close price, and trading volume at $t$-th time step, respectively. Since most of the indicators are calculated within a certain period, we denote $n$ as that time window.
\begin{itemize}
    \item Arithmetic ratio (AR): The open, high, and low price ratio over the close price.
    \begin{equation}
    \label{eq:arithmetic_ratio}
    AR_O = \frac{O_t}{C_t},
    AR_H = \frac{H_t}{C_t},
    AR_L = \frac{L_t}{C_t}
    \end{equation}
    
    \item Close Price Ratio: The ratio of close price over the highest and the lowest close price within a time window.
    \begin{equation}
        \label{eq:close_ratio}
        \begin{split}
        &RC_{min_t} = \frac{C_t}{\text{min}([C_t, C_{t-1}, ..., C_{t-n}])}\\
        &RC_{max_t} = \frac{C_t}{\text{max}([C_t, C_{t-1}, ..., C_{t-n}])}
        \end{split}
    \end{equation}
    
    \item Close SMA: The simple moving average of the close price over a time window 
    \begin{equation}
        \label{eq:close_sma}
        SMA_{C_t} = \frac{C_t + C_{t-1} + ... + C_{t-n}}{n}
    \end{equation}
    
    \item Close EMA: The exponential moving average of the close price over a time window
    \begin{equation}
        \label{eq:close_ema}
        EMA_{C_t} = C_t * k + EMA_{C_{t-1}} * (1-k)
    \end{equation}
    where: $k=\frac{2}{n+1}$
    
    
    \item Volume SMA: The simple moving average of the volume over a time window
    \begin{equation}
        \label{eq:volume_sma}
        SMA_{V_t} = \frac{V_t + V_{t-1} + ... + V_{t-n}}{n}
    \end{equation}
    
    \item Volume EMA: The exponential moving average of the close price over a time window
    \begin{equation}
        \label{eq:volume_ema}
        EMA_{V_t} = V_t * k + EMA_{V_{t-1}} * (1-k)
    \end{equation}
    where: $k=\frac{2}{n+1}$

    \item Average Directional Index (ADX): ADX is used to quantify trend strength. ADX calculations are based on a moving average of price range expansion over a given period of time.
    \begin{equation}
        \label{eq:adx}
        \begin{split}
            &DI^+_t=\frac{\text{Smoothed } DM^+}{ATR_t} * 100\\
            &DI^-_t=\frac{\text{Smoothed } DM^-}{ATR_t} * 100\\
            &DX_t=\frac{DI^+_t - DI^-_t}{DI^+_t + DI^-_t} * 100\\
            &ADX_t=\frac{ADX_{t-1}*(n-1) + DX_t}{n}
        \end{split}
    \end{equation}
    where:
    \begin{itemize}
        \item $DM^+ = H_t - H_{t-1}$
        \item $DM^- = L_{t-1} - L_t$
        \item $\text{Smoothed } DM^{+/-} = DM_{t-1} - \frac{DM_{t-1}}{n} + DM_t $
        \item ATR: Average True Range
    \end{itemize}

    \item Relative Strength Index (RSI): measures the magnitude of recent price changes to evaluate overbought or oversold conditions in the price of a stock or other asset. It is the normalized ration of the average gain over the average loss.
    
    \begin{equation}
        \label{eq:rsi}
        \begin{split}
        &avgGain_t = \frac{(n-1)*avgGain_{t-1}+gain_t}{n}\\
        &avgLoss_t = \frac{(n-1)*avgLoss_{t-1}+loss_t}{n}\\
        &RSI_t = 100 - \frac{100}{1+\frac{avgGain_t}{avgLoss_t}}
        \end{split}
    \end{equation}
    where:
    \begin{itemize}
        \item $gain_t = 
        \begin{cases}
            C_t - C_{t-1},& \text{if } C_t > C_{t-1}\\
            0,& \text{otherwise}
        \end{cases}
        $
        \item $loss_t = 
        \begin{cases}
            0,& \text{if } C_t > C_{t-1}\\
            C_{t-1} - C_t,& \text{otherwise}
        \end{cases}
        $
    \end{itemize}
    
    \item Moving average convergence divergence (MACD): shows the relationship between two moving averages of a stock’s price. It is calculated by the subtraction of the long-term EMA from the short-term EMA.
    \begin{equation}
        \label{eq:macd}
        MACD_t = EMA_t(n=12) - EMA_t(n=26)
    \end{equation}
    where:
    \begin{itemize}
        \item $EMA_t(n=12)$: The exponential moving average at $t$-th time step of the close price over 12-time steps.
        \item $EMA_t(n=26)$: The exponential moving average at $t$-th time step of the close price over 26-time steps.
    \end{itemize}

    \item Stochastics: an oscillator indicator that points to buying or selling opportunities based on momentum
    \begin{equation}
        \label{eq:stochastics}
        Stochastic_t = 100 * \frac{Cur_{t} - L_{t-n}}{H_{t-n} - L_{t-n}}
    \end{equation}
    
    where:
    \begin{itemize}
        \item $TP_t = \frac{H_t+L_t+C_t}{3}$
        \item $moneyFlow_t = TP_t * Volume_t$
        \item $posMF_t = posMF_{t-1} + moneyFlow_t \text{ if } TP_t > TP_{t-1}$
        \item $negMF_t = negMF_{t-1} + moneyFlow_t \text{ if } TP_t \leq TP_{t-1}$
    \end{itemize}

    \item Money Flow Index (MFI): an oscillator measures the flow of money into and out over a specified period of time. The MFI is the normalized ratio of accumulating positive money flow (upticks) over negative money flow values (downticks).
    \begin{equation}
        \label{eq:mfi}
        MFI_t = 100 - \frac{100}{1+\frac{\sum_{t-n}^t posMF}{\sum_{t-n}^t negMF}}
    \end{equation}
    
    where:
    \begin{itemize}
        \item $TP_t = \frac{H_t+L_t+C_t}{3}$
        \item $moneyFlow_t = TP_t * Volume_t$
        \item $posMF_t = posMF_{t-1} + moneyFlow_t \text{ if } TP_t > TP_{t-1}$
        \item $negMF_t = negMF_{t-1} + moneyFlow_t \text{ if } TP_t \leq TP_{t-1}$
    \end{itemize}
    
    \item Average of True Ranges (ATR): The simple moving average of a series of true range indicators. True range indicators show the max range between \textit{(High - Low), (High-Previous\_Close), and (Previous\_Close - Low)}.
    \begin{equation}
        \label{eq:atr}
        \begin{split}
        &TR_t = \text{Max}(H_t-L_t, |H_t-C_{t-1}|, |L_t-C_{t-1}|)\\
        &ATR_t = \frac{1}{n}\sum_{i=t-n}^{t}TR_i
        \end{split}
    \end{equation}

    \item Bollinger Band (BB): a set of trendlines plotted two standard deviations (positively and negatively) away from a simple moving average (SMA) of a stock's price.
    \begin{equation}
        \label{eq:bb}
        \begin{split}
            &BOLU_t = MA_t(TP_t, w) + m*\sigma[TP_t, w]\\
            &BOLD_t = MA_t(TP_t, w) - m*\sigma[TP_t, w]
        \end{split}
    \end{equation}
    
    where:
    \begin{itemize}
        \item $BOLU_t$: Upper Bollinger Band at $t$-th time step
        \item $BOLD_t$: Lower Bollinger Band at $t$-th time step
        \item $MA_t$: Moving average at $t$-th time step
        \item $TP_t = \frac{H_t+L_t+C_t}{3}$
        \item $m$: number of standard deviations (typically 2)
        \item $\sigma[TP, n]$: Standard deviation over last w periods of $TP$
    \end{itemize}
    
    \item On-Balance Volume (OBV): measures buying and selling pressure as a cumulative indicator that adds volume on up days and subtracts volume on down days.
    \begin{equation}
        \label{eq:obv}
        OBV_t = OBV_{t-1} + 
        \begin{cases}
            V_t,& \text{if } C_t > C_{t-1}\\
            0,& \text{if } C_t = C_{t-1}\\
            -V_t,& \text{if } C_t < C_{t-1}\\
        \end{cases}
    \end{equation}
    
    
\end{itemize}


\end{document}